\documentclass[journal]{IEEEtran}
\ifCLASSINFOpdf
\else
\fi
\usepackage[utf8]{inputenc}
\usepackage{amsmath,amssymb,amsfonts}
\usepackage{cite}

\usepackage{algorithmic}
\usepackage{stfloats}
\usepackage{graphicx}
\usepackage{textcomp}
\usepackage{xcolor}
\usepackage{color}
\usepackage{pifont}
\usepackage{amsthm}

\usepackage{hyperref}
\usepackage{mathrsfs}
\usepackage{booktabs}
\usepackage{hyperref}
\usepackage{cases}
\usepackage{comment}
\usepackage{empheq} 
\usepackage{caption}
\usepackage{subcaption}
\usepackage{extarrows}
\allowdisplaybreaks[4]
\usepackage{url}  
\usepackage{graphicx}  
\usepackage[ruled,vlined,linesnumbered]{algorithm2e}

\usepackage{listings}

\begin{document}
\captionsetup[figure]{labelfont={rm},labelformat={default},labelsep=period,name={Fig.}}
\title{{Meta-Reinforcement Learning} Optimization for Movable Antenna-aided Full-Duplex CF-DFRC Systems with Carrier Frequency Offset}
%
%
%

\author{Yue Xiu,~Wanting Lyu,~You Li,~Ran Yang,~Phee Lep Yeoh,~\IEEEmembership{Senior Member,~IEEE},\\
~Wei Zhang,~\IEEEmembership{ Fellow,~IEEE},~Guangyi Liu,~\IEEEmembership{Senior Member,~IEEE},~Ning Wei\\
\thanks{Yue Xiu, Wanting Lyu, Ran Yang and Ning Wei are with 
National Key Laboratory of Science and Technology on Communications, University of Electronic Science and Technology of China, Chengdu 611731, China (E-mail:  
xiuyue12345678@163.com,lyuwanting@yeah.net, yangran6710@outlook.com, wn@uestc.edu.cn).}
\thanks{You Li is with the Southwest China Research Institute of Electronic Equipment(SWIEE), China (e-mail:liyou1992@163.com).}
\thanks{Phee Lep Yeoh is with the School of Science, Technology and Engineering,
University of the Sunshine Coast, Sippy Downs, QLD 4556, Australia (e-mail:
pyeoh@usc.edu.au).}
\thanks{Wei Zhang is with the School of Electrical Engineering and Telecommu
nications, University of New South Wales, Sydney, NSW 2052, Australia
 (e-mail: w.zhang@unsw.edu.au).}
\thanks{G. Liu is with the China Mobile Research Institute, Beijing 100053, China
(email: liuguangyi@chinamobile.com).}
\thanks{The corresponding author is Ning Wei.}
}

\maketitle
\begin{abstract}
By enabling spectrum sharing between radar and communication operations, the cell-free dual-functional radar–communication (CF-DFRC) system {is a promising candidate to significantly improve spectrum efficiency in} future sixth-generation (6G) wireless networks. However, in wideband scenarios, synchronization errors caused by carrier frequency offset (CFO) can severely reduce both communication capacity and sensing accuracy. To address this challenge, this paper integrates movable antennas (MAs) into the CF-DFRC framework, leveraging their spatial flexibility and adaptive beamforming to dynamically mitigate CFO-induced impairments. To fully exploit the advantages of MAs in wideband scenarios with CFO, {we aim} to maximize the worst-case sum-rate of communication and sensing by jointly optimizing MA positions, {beamforming}, and CFO parameters, subject to transmit power and MA positioning constraints. Due to the non-convex nature of the {problem}, {we propose a robust meta reinforcement learning (MRL)-based} two-stage alternating optimization {strategy}. In the first stage, {we employ manifold optimization (MO) with penalty dual decomposition (PDD) to solve the CFO-robust worst-case subproblem}. In the second stage, we adopt to jointly optimize {the MA positions and beamforming vectors} in a data-driven manner {for dynamic wireless environments}. Simulation results show that the proposed {MRL} approach significantly outperforms conventional {deep reinforcement learning (DRL)} schemes in both communication and sensing performance under CFO impairments. Furthermore, compared to fixed-position antennas (FPAs), the MA-aided CF-DFRC system exhibits enhanced robustness against CFO effects.
\end{abstract}

\section{Introduction}
Dual-functional radar-communication (DFRC) has emerged as a key enabling technology for sixth-generation (6G) {wireless} applications such as smart cities, autonomous vehicles, and remote healthcare\cite{b1}. Unlike conventional communication systems, DFRC integrates sensing and communication within a unified hardware platform, thereby enhancing spectrum efficiency and enabling hardware resource sharing\cite{b2}. Meanwhile, the emerging cell-free (CF) system architecture {has been proposed to improve coverage and data rates by jointly serving user equipments (UEs) using a cooperative} network of geographically distributed {access points (APs)} interconnected {via backhaul links} with a central processing unit (CPU). To further enhance spectral efficiency while maintaining robust communication and sensing capabilities, recent research has explored integrating CF with DFRC, leading to the cell-free dual-functional radar-communication (CF-DFRC) system\cite{b3}. {By combining the benefits of distributed access and spectrum sharing, CF-DFRC improves interference management, balances network coverage, and enhances performance in high-mobility scenarios.} 

{A significant challenge in} wideband CF-DFRC {is} systems, carrier frequency offset (CFO) {that} inevitably arises from oscillator mismatches \cite{b4}. This leads to degraded synchronization accuracy, which adversely affects both communication rate and sensing accuracy. Therefore, precise frequency synchronization is crucial for enabling efficient wideband communication and high-accuracy sensing. Furthermore, limited system resources and dynamic mobility conditions intensify the impact of CFO, further undermining system robustness and overall performance. As a result, designing a robust optimization framework to mitigate CFO-induced impairments is a key challenge in realizing practical CF-DFRC deployments.

As the number of APs increases, the impact of CFO becomes more severe due to higher synchronization complexity and accumulated frequency mismatches across APs. {Recent research has considered the use of reconfigurable intelligent surfaces (RISs) to mitigate CFO-induced impairments via passive beamforming\cite{b5}}. However, the deployment of RIS {can be} constrained by environmental factors {which} motivates the need for more flexible solutions. In this context, {we propose to employ} movable antennas (MAs) {at the APs} as a promising alternative technology {for CF-DFRC systems}. Unlike traditional {fixed position antenna (FPA)} architectures, MAs allow adjustment of antenna positions \cite{b6}, enabling real-time {optimization in wireless channel} environments. By adaptively adjusting their positions and orientations based on channel state information (CSI), MAs can mitigate path loss, suppress interference, and enhance spatial degrees of freedom (DoFs). This significantly enhances both communication and sensing performance, making MAs well-suited for robust signal transmission in CF-DFRC with CFO{impairments.}

A key challenge {of deploying MAs} lies in accurately modeling the effect of CFO on sensing accuracy, particularly in wideband scenarios. To address this issue, this paper proposes {a robust optimization framefore for} MA-aided CF-DFRC, where multiple distributed APs jointly optimize beamforming and MA positions to enable reliable communication and accurate sensing.
The main contributions of this work are summarized as follows.
\begin{itemize}
    \item First, we derive the worst-case weighted communication and sensing rate (WCSR) for target position estimation under CFO conditions. The results demonstrate that the CFO increases the {Cramer-Rao lower bound (CRLB)} of target position estimation, thereby {decreasing} sensing accuracy. Based on this analysis, we propose a robust optimization framework for the wideband MA-aided CF-DFRC system, aiming to maximize the worst-case WCSR. To solve the resulting worst-case CFO optimization problem, we develop an {alternating} optimization (AO) algorithm that leverages manifold optimization (MO) and penalty dual decomposition (PDD) methods. This approach effectively addresses the robustness challenges caused by the CFO and offers a novel solution for enhancing CF-DFRC system performance under CFO.
    \item To solve the joint beamforming and MA position optimization problem under CFO, we propose a meta-reinforcement learning (MRL) framework capable of quickly adapting to dynamic {wireless channel} environments. Unlike conventional optimization methods that are typically tailored to static or specific scenarios, the proposed MRL framework leverages a meta-training phase to enable fast adaptation across heterogeneous tasks and environmental variations. This significantly improves learning efficiency and convergence speed in complex multi-user settings. Moreover, the MRL-based framework demonstrates superior scalability and flexibility, making it well-suited for real-time optimization in MA-aided CF-DFRC systems.
    \item Simulation results demonstrate that the proposed AO and MRL based robust framework significantly improves the resilience of CF-DFRC systems against CFO-induced impairments. Compared with conventional deep reinforcement learning (DRL) approaches, {our} framework achieves {better} performance improvement {with faster} convergence time, benefiting from its adaptive learning and optimization capability under varying CFO conditions.
    In addition, simulation results reveal that MAs consistently outperform fixed-position antennas (FPAs) in dynamic environments. These results validate the effectiveness of the proposed framework in enhancing both communication and sensing performance, offering a robust solution for wideband CF-DFRC systems.
\end{itemize}
\section{Relative Work}
\subsection{Resource Allocation Over CF-DFRC} {Recently, intensive} research has been conducted on CF-DFRC resource allocation to enhance overall system performance and quality of service (QoS). A communication-sensing region was introduced in \cite{b7}, and corresponding resource schedule strategies were developed to enable coordinated service coverage. In \cite{b8}, a multi-APs cooperative target detection scenario was investigated, where a joint power allocation and sensing strategy was proposed to minimize detection errors while maintaining communication quality, highlighting the importance of resource optimization for sensing reliability.
Security-aware resource allocation was studied in \cite{b9}, where the presence of both communication and sensing eavesdroppers was considered. A joint optimization framework was proposed to maximize the secrecy rate while ensuring a sufficient sensing signal-to-noise ratio (SNR), thus achieving a performance trade-off in secure environments. In \cite{b10}, non-orthogonal multiple access (NOMA) was integrated into CF-DFRC systems, and a joint user pairing and beamforming strategy was developed to improve spectral efficiency and sensing performance.
A user-centric design approach was investigated in \cite{b11}, proposing a joint resource allocation scheme that considers the sensing accuracy constraints. The scheme significantly improved overall system efficiency by dynamically allocating frequency, power, and time resources. In \cite{b12}, a multi-user optimization model was formulated under secrecy and sensing constraints, and an associated allocation strategy was developed to enhance system robustness and confidentiality.
In multi-antenna CF-DFRC systems, \cite{b13} focused on joint beamforming design {for FPA architectures} by formulating a multi-objective optimization problem, enabling cooperative enhancement of sensing and communication links.

\subsection{DFRC System With CFO}
{Robust} techniques to mitigate CFO {in DFRC systems} have obtained considerable {research} attention. {In}\cite{b14}{, the authors} proposed a cooperative passive sensing method based on mobile communication systems, where CFO effects are incorporated into the system design{, showing that o}ptimizing signal synchronization enhances both sensing and communication performance in dynamic environments. \cite{b15} addressed CFO estimation in asynchronous mobile devices, focusing on bistatic Doppler frequency estimation. A novel frequency estimation method was introduced that improves synchronization precision and minimizes the negative impact of CFO on system performance. In\cite{b16}, a fingerprint-spectrum-based synchronization method was presented to effectively resolve synchronization errors induced by CFO in asynchronous sensing mobile networks, thereby improving time-frequency synchronization and network robustness in multi-user environments.
Moreover, \cite{b17} examined clutter suppression, time-frequency synchronization, and sensing parameter association in asynchronous vehicular networks. A robust synchronization algorithm was proposed to counteract time-frequency shifts caused by CFO, aiming to enhance sensing accuracy and communication reliability in vehicular networks. \cite{b18} investigated uplink sensing in asynchronous sensing mobile networks, proposing a synchronization algorithm that integrates CFO compensation. Optimizing beamforming improves the system's sensing capability and reduces the impact of CFO on localization accuracy. \cite{b19} introduced RIS and explored user localization and environment mapping with RIS assistance. A new synchronization mechanism was developed that ensures localization accuracy even under CFO, thereby significantly improving the robustness of DFRC systems against CFO.

\subsection{MA-Aided DFRC System}
{The introduction of} MAs has opened new pathways for enhancing system flexibility and performance {in wireless systems}. Traditional FPAs often encounter challenges such as limited coverage and suboptimal resource utilization, especially in dynamic environments. In contrast, MA structures enable the antenna positions, thereby facilitating joint improvements in communication and sensing rates. As a result, MA-aided DFRC systems have become a promising technology. {Several} studies have addressed the resource allocation challenges within MA-aided DFRC systems to develop robust and adaptive optimization strategies.
\cite{b6} introduced the concept of dynamic radar cross section (RCS) modeling and proposed a method to dynamically optimize sensing beam directions based on target characteristics, thereby providing resource allocation strategies. \cite{b20} addressed beamforming flexibility in MA-aided systems and designed a joint beamforming scheme to support communication and sensing tasks simultaneously. In\cite{b21}, the antenna positions and transmit signal were jointly modeled under the MA architecture, demonstrating that optimizing antenna positions plays a crucial role in resource scheduling by enhancing both sensing accuracy and communication reliability. For integrated air-ground DFRC scenarios supporting the low-altitude economy, \cite{b22} proposed an MA-DFRC framework that emphasizes the impact of antenna movement on spatial partitioning and resource prioritization and developed a resource matching mechanism that balances sensing coverage with communication quality.
Building upon these contributions, \cite{b23} incorporated RIS into MA-DFRC systems and explored joint antenna and RIS control for spatially selective resource allocation, leading to improved spectral efficiency.  \cite{b24} proposed a joint beamforming model for MA-based aerial platforms, integrating power allocation into the beamforming process to establish a closed-loop link between channel states and resource configurations. From a performance bounds perspective, \cite{b25} formulated a Cramér-Rao bound minimization approach to design a resource allocation method for multi-user MA-DFRC systems, facilitating collaborative user scheduling and weighted resource assignment, thus simultaneously enhancing sensing precision and communication capacity. 

\section{System Model}
\begin{figure}[htbp]
  \centering
  \includegraphics[width=0.45\textwidth, height=0.5\textwidth]{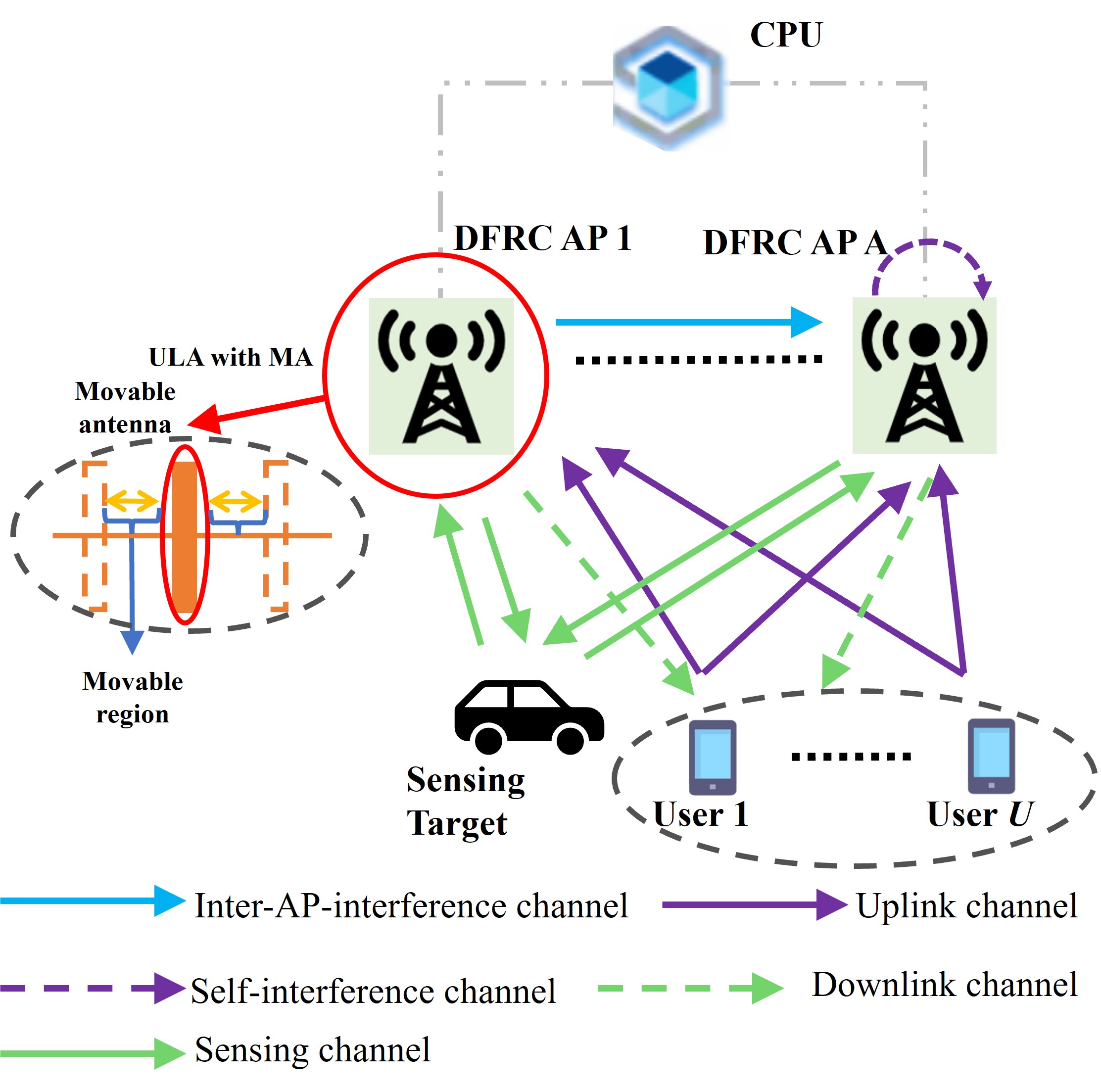}
  \captionsetup{justification=centering}
  \caption{Illustration of the MA-aided CF-DFRC system.}
\label{FIGURECFO1}
\end{figure}

\subsection{System composition}
In this paper, we investigate a full-duplex (FD) CF-DFRC system enhanced by MAs. The system comprises $A$ DFRC APs, where each DFRC AP is equipped with $N$ transmit MAs, and $M$ receive MAs. These APs collaboratively serve $U$ single-antenna UEs while performing target sensing tasks, as illustrated in Fig.\ref{FIGURECFO1}. Without loss of generality, we denote the sets of UEs and DFRC APs by $\mathcal{U}=\{1,\ldots,U\}$ and $\mathcal{A}=\{1,\ldots,A\}$, respectively. The position vector of the transmit MA of the $a$th AP is denoted as $\mathbf{t}_{a}=[t_{1}^{a},\ldots,t_{N}^{a}]$, and the position vector of the receive MA of the $a$th AP is given by $\mathbf{r}_{a}=[r_{1}^{a},\ldots,r_{M}^{a}]$. We assume that the CSI between each UE and all APs is perfectly known\cite{b20}.

To account for potential synchronization imperfections, we consider the CFO that may exist between different DFRC APs. Although each AP is assumed to achieve full synchronization of its transmit MAs, either through the Argos algorithm proposed in \cite{b26} or by linking all transmit MAs to a common local oscillator, such synchronization only ensures intra-AP alignment. As pointed out in \cite{b4,b27},  residual CFO may still occur across APs due to hardware imperfections {or} independent oscillators. Therefore, we denote the CFO between the $a$th and the $a^{\prime}$th APs as $\Delta f_{a,a^{\prime}}$. According to the discussion in\cite{b28}, the Orthogonal frequency-division multiplexing (OFDM) signal in the frequency domain consists of multiple subcarriers, and the CFO affects each subcarrier. As a result, the CFO model for the OFDM symbol can be represented as 
\begin{align}
c_{a,a^{\prime}}[s]=x_{a}[s]e^{j2\pi\Delta f_{a,a^{\prime}}T_{sym}},\forall~a^{\prime}\neq a, \label{pro1}
\end{align}    
where $T_{sym}$ denotes the duration of each OFDM symbol, and $x_{a}[s]$ represents the OFDM symbol on the $s$th subcarrier{, where} $s\in\mathcal{S}=\{1,\ldots,S\}$, and $\mathcal{S}$ denotes the set of subcarriers.

\subsection{Channel model}
This paper considers {a} multipath channel under quasi-static block-fading conditions, where channel parameters remain constant within the duration of a single OFDM symbol \cite{b29}. {Referring to} Fig.~\ref{FIGURECFO1}, the elevation angle of departure (AoD) {for each wireless channel} is defined as $\theta_{AoD} \in \left[-\frac{\pi}{2}, \frac{\pi}{2}\right]$. Building on the field-response theory proposed in \cite{b30}, we develop a channel model tailored to MA-aided systems. This framework includes the following {channel models:} 
\begin{itemize}
     \item Uplink channel $\mathbf{h}_{u}(\mathbf{r}_{a})$,
    \item Self-interference (SI) channel $\mathbf{H}_{SI}(\mathbf{t}_{a},\mathbf{r}_{a})$,
    \item Inter-AP-interference (IAI) channel $\mathbf{G}(\mathbf{t}_{a},\mathbf{r}_{a^{\prime}})$,
    
    \item Sensing channel $\mathbf{H}_{T}(\mathbf{t}_{a},\mathbf{r}_{a})$.
\end{itemize}
The details of the channel models are given as follows.

\textbf{(1) Uplink channels:}
In this paper, all users are equipped with a single antenna, so the field response matrix (FRM) exists only at the uplink receive AP. 
Specifically, let 
$L_{u,a}$ represents the number of uplink paths from the 
$u$th user to the $a$th AP. 
Based on these parameters, the system's uplink channel transmission models can be represented as, respectively 
\begin{align}
&\mathbf{h}_{u}(\mathbf{r}_{a})=\bar{\mathbf{h}}_{u}^{H}\mathbf{F}_{up}(\mathbf{t}_{a}),\label{pro2}
\end{align}
where 
$\mathbf{F}_{up}(\mathbf{t}_{a})\in\mathbb{C}^{L_{u,a}\times N}$ represents the receiver FRM of the uplink channel. The channel gain vectors 
$\bar{\mathbf{h}}_{u}\in\mathbb{C}^{L_{u,a}\times 1}$ is defined as the uplink channel response characteristics from user $u$ to the $a$th AP.

\textbf{(2) SI and IAI channels:}
Based on the definition in equation (\ref{pro2}), we continue to model the SI and IAI channels using field response theory. Since both the receiver and transmitter of the SI and IAI channels are equipped with multiple antennas, the SI and IAI channels are modeled as 
\begin{align}
&\mathbf{H}(\mathbf{t}_{a},\mathbf{r}_{a})=\bar{\mathbf{F}}(\mathbf{r}_{a})^{H}\mathbf{\Sigma}_{SI,a}\bar{\mathbf{G}}(\mathbf{t}_{a}),\nonumber\\
&\mathbf{G}(\mathbf{t}_{a},\mathbf{r}_{b})=\tilde{\mathbf{F}}(\mathbf{r}_{a^{\prime}})^{H}\mathbf{\Sigma}_{IAI,a,a^{\prime}}\tilde{\mathbf{G}}(\mathbf{t}_{a}),\label{pro3}
\end{align}
in which
$\bar{\mathbf{F}}(\mathbf{r}_{a})\in\mathbb{C}^{L_{SI,a}\times M}$ and 
$\bar{\mathbf{G}}(\mathbf{t}_{a})\in\mathbb{C}^{L_{SI,a}\times N}$ are the transmitter and receiver FRMs of the SI channel, while 
$\tilde{\mathbf{F}}(\mathbf{r}_{a^{\prime}})\in\mathbb{C}^{L_{IAI,b}\times M}$ and 
$\tilde{\mathbf{G}}(\mathbf{t}_{a})\in\mathbb{C}^{L_{IAI,a}\times N}$ represent the transmitter and receiver FRMs of {the} IAI channel. $L_{SI,a}$ and 
$L_{IAI,a}$ denote the number of transmit paths for the SI and IAI channels, respectively, and 
$\bar{L}_{SI,a}$ and $L_{IAI,a^{\prime}}$ are the number of receive paths for the SI and IAI channels, respectively{. Furthermore,} $\mathbf{\Sigma}_{SI,a}\in\mathbb{C}^{L_{SI,a}\times L_{SI,a}}$ and $\mathbf{\Sigma}_{IAI,a,a^{\prime}}\in\mathbb{C}^{L_{IAI,a^{\prime}}\times L_{SI,a}}$ denote the channel gain of SI and IAI channels, respectively.

\textbf{(3) Sensing channels:}
Based on the definition in equation (\ref{pro3}), we continue to model the sensing channels based on field response theory \cite{b30}. {According to} the sensing channel model proposed in \cite{b31}, we assume {the} sensing channel follows a line of sight (LoS) model.
Since both the receiver and transmitter of the sensing channels are equipped with multiple antennas, sensing channels are modeled as 
\begin{align}
&\mathbf{H}_{T}(\mathbf{t}_{a},\mathbf{r}_{a})=\rho_{T}\mathbf{f}_{T}(\mathbf{r}_{a})\mathbf{g}_{T}^{H}(\mathbf{t}_{a}),\label{pro4}
\end{align}
in which
$\mathbf{g}_{T}(\mathbf{t}_{a})\in\mathbb{C}^{N\times 1}$ and 
$\mathbf{f}_{T}(\mathbf{r}_{a})\in\mathbb{C}^{M\times 1}$ are the transmitter and receiver FRMs of the sensing channel{, respectively, while} $\rho_{T}$ denotes the RCS.

\textbf{(4) FRM model:}
In this section, we introduce the FRM model of SI channel.  
Specifically, the difference of the signal propagation distance for the $l$th $(1\leq l\leq L_{SI,a})$
transmit path between the MA position $\mathbf{t}_{a}$
and the origin
of the transmit region, i.e., $O_{a}$ in Fig.\ref{FIGURECFO1}, can be expressed
as $\rho_{SI,l}^{a}(t_{n}^{a})=t_{n}^{a}\cos\theta_{SI,l}^{a}$, where $\lambda$ is the carrier wavelength, the phase difference is
calculated by $\frac{2\pi}{\lambda}\rho_{SI,l}^{a}(t_{n}^{a})$. Thus, the transmit FRM, which characterizes the phase differences of $L_{SI,a}$
transmit paths, are denoted as
\begin{align}
\mathbf{g}(t_{n}^{a})=\left[e^{j\frac{2\pi}{\lambda}\rho_{SI,1}^{a}(t_{n}^{a})},\ldots,e^{j\frac{2\pi}{\lambda}\rho_{SI,L_{SI,a}}^{a}(t_{n}^{a})}\right]^{T}\in\mathbb{C}^{L_{SI,a}\times 1}.\label{pro5}
\end{align}
Similarly, the receive FRM is given by
\begin{align}
\mathbf{f}(r_{m}^{a})=\left[e^{j\frac{2\pi}{\lambda}\rho_{SI,1}^{a}(r_{m}^{a})},\ldots,e^{j\frac{2\pi}{\lambda}\rho_{SI,L_{SI}^{a}}^{a}(r_{m}^{a})}\right]^{T}\in\mathbb{C}^{\bar{L}_{SI,a}\times 1}.\label{pro6}
\end{align}
which represents the phase differences of $\bar{L}_{SI,a}$ receive paths,
where $\rho_{SI,l^{a}}^{a}(r_{m}^{a})=r_{m}^{a}\cos\theta_{SI,l_{2}}^{a}(1\leq l_{2}\leq \bar{L}_{SI,a})$ is the difference of the signal propagation
distance for the $l_{2}$th receive path between the MA position
$r_{m}^{a}$
and the origin of the receive region. Thus, 
$\bar{\mathbf{F}}(\mathbf{r}_{a})$ and $\bar{\mathbf{G}}(\mathbf{t}_{a})$
are given by
\begin{align}
&\bar{\mathbf{F}}(\mathbf{r}_{a})=[\mathbf{f}(r_{1}^{a}),\ldots,\mathbf{f}(r_{M}^{a})],\nonumber\\
&\bar{\mathbf{G}}(\mathbf{t}_{a})=[\mathbf{g}(t_{1}^{a}),\ldots,\mathbf{g}(t_{N}^{a})].\label{pro7}
\end{align}
Since other FRMs
have the
similar structures as $\bar{\mathbf{F}}(\mathbf{r}_{a})$ and $\bar{\mathbf{G}}(\mathbf{t}_{a})$
, we can obtain the other FRMs from the uplink, downlink, IAI and sensing channels by replacing the {corresponding} index {set}.

\subsection{Received Signal Model}
The received signal at the $a$th AP includes the uplink signals from the $U$ communication users, the reflected echo, and the interference from APs $a^{\prime}$, the IAI, as well as the SI, which can be represented as 
\begin{align}
&\mathbf{y}_{a}[s]=\underbrace{\sum\nolimits_{u=1}^{U}\mathbf{h}_{u}(\mathbf{t}_{a})\sqrt{p_{u}}x_{u}}_{\text{Uplink communication signal}}+
\underbrace{\mathbf{H}(\mathbf{t}_{a},\mathbf{r}_{a})\mathbf{w}_{a}x_{a}[s]}_{\text{SI signal}}\nonumber\\
&+\underbrace{\mathbf{H}_{T}(\mathbf{t}_{a},\mathbf{r}_{a})\mathbf{w}_{a}x_{a}[s]}_{\text{Sensing signal}}+\underbrace{\sum\nolimits_{a^{\prime}=1,\mathcal{A}\setminus \{a\} }^{A}\mathbf{H}_{T}(\mathbf{t}_{a^{\prime}},\mathbf{r}_{a})\mathbf{w}_{a^{\prime}}c_{a,a^{\prime}}[s]}_{\text{Cross sensing signal}}\nonumber\\
&+\underbrace{\sum\nolimits_{a^{\prime}=1,\mathcal{A}\setminus \{a\} }^{A}\mathbf{G}(\mathbf{t}_{a},\mathbf{r}_{a^{\prime}})\mathbf{w}_{a^{\prime}}c_{a,a^{\prime}}[s]}_{\text{IAI signal}}+n_{a}[s],\label{pro_8}
\end{align}
where $\mathcal{A}\setminus\{a\}$ represents the set 
$\mathcal{A}$ with the index $a$ removed.
Since the SI and sensing signals are typically assumed to share the same local oscillator as the transmitter, the CFO is omitted in the corresponding SI and sensing terms in (\ref{pro_8}), {where the cross sensing signal refers to the unintended leakage of radar echoes or communication signals between distributed units, which can be exploited for cooperative sensing; however, CFO introduces phase misalignment and frequency mismatch, which distort the cross sensing signals and exacerbate IAI, thereby degrading both sensing and communication accuracy.} Therefore, the transmit signal defined in (\ref{pro1}) is substituted into (\ref{pro_8}), which is reformulated as 
\begin{align}
&\mathbf{y}_{a}[s]=\sum\nolimits_{u=1}^{U}\mathbf{h}_{u}(\mathbf{t}_{a})\sqrt{p_{u}}x_{u}+
\mathbf{H}(\mathbf{t}_{a},\mathbf{r}_{a})\mathbf{w}_{a}x_{a}[s]\nonumber\\
&+\mathbf{H}_{T}(\mathbf{t}_{a},\mathbf{r}_{a})\mathbf{w}_{a}x_{a}[s]\nonumber\\
&+\sum\nolimits_{a^{\prime}=1,\mathcal{A}\setminus \{a\} }^{A}\mathbf{H}_{T}(\mathbf{t}_{a^{\prime}},\mathbf{r}_{a})\mathbf{w}_{a^{\prime}}e^{j2\pi s\Delta f_{a,a^{\prime}}T_{sym}}x_{a^{\prime}}[s]\nonumber\\
&+\sum\nolimits_{a^{\prime}=1,\mathcal{A}\setminus \{a\} }^{A}\mathbf{G}(\mathbf{t}_{a},\mathbf{r}_{a^{\prime}})\mathbf{w}_{a^{\prime}}e^{j2\pi s\Delta f_{a,a^{\prime}}T_{sym}}x_{a^{\prime}}[s]+n_{a}[s].\label{pro8}
\end{align}
Next, the CPU collects signals from $S$ subcarriers into {a} vector $SM\times 1$. The received signal at the $a$th AP is defined as $\mathbf{y}_{a}\in\mathbb{C}^{SM\times 1}$, {which can be expressed as}
\begin{align}
&\mathbf{y}_{a}=\underbrace{\mathbf{1}_{S\times 1}\otimes\left(\sum\nolimits_{u=1}^{U}\mathbf{h}_{u}(\mathbf{t}_{a})\sqrt{p_{u}}x_{u}\right)}_{\mathbf{\Xi}(\mathbf{t}_{a})}+\underbrace{\mathbf{X}_{a}\mathbf{H}(\mathbf{t}_{a},\mathbf{r}_{a})\mathbf{w}_{a}}_{\mathbf{B}(\mathbf{w}_{a},\mathbf{t}_{a},\mathbf{r}_{a})}\nonumber\\
&+\underbrace{\mathbf{X}_{a}\mathbf{H}_{T}(\mathbf{t}_{a},\mathbf{r}_{a})\mathbf{w}_{a}}_{\mathbf{D}(\mathbf{w}_{a},\mathbf{t}_{a},\mathbf{r}_{a})}\nonumber\\
&+\sum\nolimits_{a^{\prime}=1,\mathcal{A}\setminus \{a\} }^{A}\underbrace{\mathrm{diag}\{\mathbf{X}_{a^{\prime}}\mathbf{H}_{T}(\mathbf{t}_{a^{\prime}},\mathbf{r}_{a})\mathbf{w}_{a^{\prime}}\}}_{\mathbf{D}(\mathbf{r}_{a},\mathbf{t}_{a^{\prime}},\mathbf{w}_{a^{\prime}})}\mathbf{u}_{a,a^{\prime}}\nonumber\\
&+\sum\nolimits_{a^{\prime}=1,\mathcal{A}\setminus \{a\} }^{A}\underbrace{\mathrm{diag}\{\mathbf{X}_{a^{\prime}}\mathbf{G}(\mathbf{t}_{a},\mathbf{r}_{a^{\prime}})\mathbf{w}_{a^{\prime}}\}}_{\mathbf{E}(\mathbf{t}_{a},\mathbf{r}_{a^{\prime}},\mathbf{w}_{a^{\prime}})}\mathbf{u}_{a,a^{\prime}}+\mathbf{n}_{a},\label{pro9}
\end{align}
where transmit signal $\mathbf{X}_{a}$ {is} expressed as
$\mathbf{X}_{a}=[x_{a}[1]\mathbf{I}_{M\times M},\ldots,x_{a}[S]\mathbf{I}_{M\times M}]^{T}\in\mathbb{C}^{MS\times M}$, $\mathbf{u}_{a,a^{\prime}}=[e^{j2\pi 1\Delta f_{a,a^{\prime}} T_{sym}},\ldots,e^{j2\pi S\Delta f_{a,a^{\prime}} T_{sym}}]^{T}\otimes\mathbf{1}_{M}$ and $\mathbf{y}_{a}=[\mathbf{y}_{a}^{T}[1],\ldots,\mathbf{y}_{a}^{T}[S]]^{T}\in\mathbb{C}^{MS\times 1}$. 
Then, $\mathbf{y}_{a}$ can be further expressed as
\begin{align}
&\mathbf{y}_{a}=\mathbf{\Xi}(\mathbf{t}_{a})+\mathbf{B}(\mathbf{w}_{a},\mathbf{t}_{a},\mathbf{r}_{a})+\mathbf{D}(\mathbf{w}_{a},\mathbf{t}_{a},\mathbf{r}_{a})+\nonumber\\
&\mathbf{D}(\mathbf{r}_{a},\mathbf{t}_{\mathcal{A}\setminus \{a\}},\mathbf{w}_{\mathcal{A}\setminus \{a\}})\mathbf{u}_{\mathcal{A}\setminus \{a\}}+\mathbf{E}(\mathbf{t}_{a},\mathbf{r}_{\mathcal{A}\setminus \{a\}},\mathbf{w}_{\mathcal{A}\setminus \{a\}})\nonumber\\
&\times\mathbf{u}_{\mathcal{A}\setminus \{a\}},\label{pro10}
\end{align}
where
\begin{align}
&\mathbf{D}(\mathbf{r}_{a},\mathbf{t}_{\mathcal{A}\setminus \{a\}},\mathbf{w}_{\mathcal{A}\setminus \{a\}})=[\mathbf{D}(\mathbf{r}_{a},\mathbf{t}_{1},\mathbf{w}_{1}),\ldots,\mathbf{D}(\mathbf{r}_{a},\mathbf{t}_{A},\mathbf{w}_{A})]\nonumber\\
&\in\mathbb{C}^{MS\times (A-1)MS},\nonumber\\
&\mathbf{E}(\mathbf{r}_{a},\mathbf{t}_{\mathcal{A}\setminus \{a\}},\mathbf{w}_{\mathcal{A}\setminus \{a\}})=[\mathbf{E}(\mathbf{r}_{a},\mathbf{t}_{1},\mathbf{w}_{1}),\ldots,\mathbf{E}(\mathbf{r}_{a},\mathbf{t}_{A},\mathbf{w}_{A})]\nonumber\\
&\in\mathbb{C}^{MS\times (A-1)MS},\nonumber\\
&\mathbf{u}_{\mathcal{A}\setminus \{a\}}=[\mathbf{u}_{1}^{T},\ldots,\mathbf{u}_{\mathcal{A}}^{T}]^{T}\in\mathbb{C}^{MS(A-1)\times 1}.
\end{align}
Then, we let $\mathbf{y}=[\mathbf{y}_{1}^{T},\ldots,\mathbf{y}_{A}^{T}]^{T}\in\mathbb{C}^{AMS\times 1}$ be the vector of the aggregated uplink signals of the $A$ APs, which are processed at the CPU. This vector can be expressed as 
\begin{align}
&\mathbf{y}=\bar{\mathbf{\Xi}}(\mathbf{r}_{\mathcal{A}})+\bar{\mathbf{B}}(\mathbf{w}_{\mathcal{A}},\mathbf{t}_{\mathcal{A}},\mathbf{r}_{\mathcal{A}})+\bar{\mathbf{D}}(\mathbf{w}_{\mathcal{A}},\mathbf{t}_{\mathcal{A}},\mathbf{r}_{\mathcal{A}})+\nonumber\\
&\tilde{\mathbf{D}}(\mathbf{r}_{\mathcal{A}},\mathbf{t}_{\mathcal{A}},\mathbf{w}_{\mathcal{A}})\mathbf{u}_{\mathcal{A}}+\tilde{\mathbf{E}}(\mathbf{t}_{\mathcal{A}},\mathbf{r}_{\mathcal{A}},\mathbf{w}_{\mathcal{A}})\mathbf{u}_{\mathcal{A}},\label{pro11}
\end{align}
where 
\begin{align}
&\mathbf{u}_{\mathcal{A}}=[\mathbf{u}_{\mathcal{A}/1}^{T},\ldots,\mathbf{u}_{\mathcal{A}\setminus \{a\}}^{T}]^{T}\in\mathbb{C}^{AMS\times 1},\nonumber\\
&\bar{\mathbf{\Xi}}(\mathbf{r}_{\mathcal{A}})=[\bar{\mathbf{\Xi}}^{T}(\mathbf{r}_{1}),\ldots,\bar{\mathbf{\Xi}}^{T}(\mathbf{r}_{A})]^{T}\in\mathbb{C}^{AMS\times 1}, \nonumber\\
&\bar{\mathbf{B}}(\mathbf{w}_{\mathcal{A}},\mathbf{t}_{\mathcal{A}},\mathbf{r}_{\mathcal{A}})=[\bar{\mathbf{B}}^{T}(\mathbf{w}_{1},\mathbf{t}_{1},\mathbf{r}_{1}),\ldots,\bar{\mathbf{B}}^{T}(\mathbf{w}_{A},\mathbf{t}_{A},\mathbf{r}_{A})]^{T}\nonumber\\
&\in\mathbb{C}^{AMS\times 1},\nonumber\\
&\bar{\mathbf{D}}(\mathbf{w}_{\mathcal{A}},\mathbf{t}_{\mathcal{A}},\mathbf{r}_{\mathcal{A}})=[\bar{\mathbf{D}}^{T}(\mathbf{w}_{1},\mathbf{t}_{1},\mathbf{r}_{1}),\ldots,\bar{\mathbf{D}}^{T}(\mathbf{w}_{A},\mathbf{t}_{A},\mathbf{r}_{A})]^{T}\nonumber\\
&\in\mathbb{C}^{AMS\times 1},\nonumber\\
&\tilde{\mathbf{D}}(\mathbf{w}_{\mathcal{A}},\mathbf{t}_{\mathcal{A}},\mathbf{r}_{\mathcal{A}})=[\mathbf{D}^{T}(\mathbf{r}_{1},\mathbf{t}_{\mathcal{A}/1},\mathbf{w}_{\mathcal{A}/1}),\ldots,\mathbf{D}^{T}(\mathbf{r}_{A},\mathbf{t}_{\mathcal{A}\setminus \{a\}}\nonumber\\
&,\mathbf{w}_{\mathcal{A}\setminus \{a\}})]^{T}\in\mathbb{C}^{AMS\times AMS},\nonumber\\
&\tilde{\mathbf{E}}(\mathbf{w}_{\mathcal{A}},\mathbf{t}_{\mathcal{A}},\mathbf{r}_{\mathcal{A}})=
[\mathbf{E}^{T}(\mathbf{t}_{1},\mathbf{r}_{\mathcal{A}/1},\mathbf{w}_{\mathcal{A}/1}),\ldots,\mathbf{E}^{T}(\mathbf{t}_{A}\nonumber\\
&,\mathbf{r}_{\mathcal{A}\setminus \{a\}},\mathbf{w}_{\mathcal{A}\setminus \{a\}})]^{T}\in\mathbb{C}^{AMS\times AMS}.
\end{align}
Since each AP is performing both downlink sensing and uplink
communication, it needs to separate the received signals into
two components: the signals intended for communication and
the signals reflected from the target. This separation can be achieved through signal processing techniques such
as filtering, beamforming, or spatial processing. Each component of the received signal (i.e., communication and echo
signals) requires different processing to optimize performance.
Receive filters are applied to each component independently to
enhance the desired signal and suppress interference or noise.
In addition, we assume that the echoes received from uplink
users do not contribute to the received signal,
given the typically lower power of uplink signals compared to downlink sensing signals. In summary, the vector in (\ref{pro11}) {can be} post-processed by the {an uplink receive beamforming} $\hat{\mathbf{u}}_{u}=[\mathbf{u}_{1,u}^{T},\ldots,\mathbf{u}_{A,U}^{T}]^{T}\in\mathbb{C}^{AM\times 1}$ to detect the signal of user $u$ that is received at the $A$ APs. Hence, the communication signal to interference plus noise ratio (SINR) of user $u$ is given by
\begin{align}
\gamma_{u}^{c}=\frac{p_{u}S|\hat{\mathbf{u}}_{u}^{H}\mathbf{h}_{u}|^{2}}{\sum_{u^{\prime}\neq u}^{U}p_{u^{\prime}}S|\hat{\mathbf{u}}_{u}^{H}\mathbf{h}_{u^{\prime}}|^{2}+I_{k}+\sigma^{2}\|\hat{\mathbf{u}}_{k}\|^{2}},\label{pro13}
\end{align}    
where 
\begin{align}
&I_{k}=|\hat{\mathbf{u}}_{k}^{H}\bar{\mathbf{B}}(\mathbf{w}_{\mathcal{A}},\mathbf{t}_{\mathcal{A}},\mathbf{r}_{\mathcal{A}})|^{2}+|\hat{\mathbf{u}}_{k}^{H}\bar{\mathbf{D}}(\mathbf{w}_{\mathcal{A}},\mathbf{t}_{\mathcal{A}},\mathbf{r}_{\mathcal{A}})|^{2}+\nonumber\\
&|\hat{\mathbf{u}}_{k}^{H}\tilde{\mathbf{D}}(\mathbf{r}_{\mathcal{A}},\mathbf{t}_{\mathcal{A}},\mathbf{w}_{\mathcal{A}})\mathbf{u}_{\mathcal{A}}|^{2}+|\hat{\mathbf{u}}_{k}^{H}\tilde{\mathbf{E}}(\mathbf{t}_{\mathcal{A}},\mathbf{r}_{\mathcal{A}},\mathbf{w}_{\mathcal{A}})\mathbf{u}_{\mathcal{A}}|^{2}.
\end{align}
Since the target detection probability relies on the target
SINR, we adopt a cooperative processing approach for the received SINRs of the $A$ APs to enhance target detection. In particular, we apply {an echo receiver} filter $\mathbf{z}=[\mathbf{z}_{1}^{T},\mathbf{z}_{2}^{T},\ldots,\mathbf{z}_{A}^{T}]^{T}\in\mathbb{C}^{AM\times 1}$
, which results in the following output radar SINR at the CPU
\begin{align}
\gamma^{r}=\frac{|\mathbf{z}\bar{\mathbf{B}}(\mathbf{w}_{\mathcal{A}},\mathbf{t}_{\mathcal{A}},\mathbf{r}_{\mathcal{A}})|^{2}}{I_{r}+\sigma^{2}\|\mathbf{z}\|^{2}},\label{pro14}
\end{align}    
where 
\begin{align}
&I_{r}=|\mathbf{z}\bar{\mathbf{\Xi}}(\mathbf{r}_{\mathcal{A}})|^{2}+|\mathbf{z}\bar{\mathbf{D}}(\mathbf{w}_{\mathcal{A}},\mathbf{t}_{\mathcal{A}},\mathbf{r}_{\mathcal{A}})|^{2}+|\mathbf{z}\tilde{\mathbf{D}}(\mathbf{r}_{\mathcal{A}},\mathbf{t}_{\mathcal{A}},\mathbf{w}_{\mathcal{A}})\mathbf{u}_{\mathcal{A}}|^{2}\nonumber\\
&+|\mathbf{z}\tilde{\mathbf{E}}(\mathbf{t}_{\mathcal{A}},\mathbf{r}_{\mathcal{A}},\mathbf{w}_{\mathcal{A}})\mathbf{u}_{\mathcal{A}}|^{2}.  
\end{align}

\subsection{Problem formulation}
This section focuses on optimizing the cooperative sensing and uplink communication of multiple APs in a MA-aided {full duplex CF-DFRC} system. Similar to the trade-off designs in \cite{b32,b33}, we aim to maximize the weighted sum-rate of sensing and communication, subject to transmit power constraints and MA position constraints. To achieve robustness against CFO, we first determine the worst-case weighted sum-rate over the CFO vector $\mathbf{u}_{\mathcal{A}}$. Subsequently, we jointly optimize the transmit beamforming $\mathbf{w}_{a}$, echo receiver filter 
$\mathbf{z}$, transmit MA position 
$\mathbf{t}_{a}$, receive MA position 
$\mathbf{r}_{a}$, uplink transmit power 
$p_{u}$, and uplink receive beamforming $\mathbf{u}_{u}$. The resulting optimization problem is formulated {as a weighted communication
and sensing rate (WCSR) expression given by}
\begin{subequations}
\begin{align}
\max_{\mathbf{z},\{p_{u}\},\{\mathbf{u}_{u}\},\atop\{\mathbf{w}_{a}\},\{\mathbf{t}_{a}\},\{\mathbf{r}_{a}\}}\min_{\mathbf{u}_{\mathcal{A}},\Delta f_{a,a^{\prime}}}&~\beta\gamma^{r}+(1-\beta)\sum\nolimits_{u=1}^{U}\gamma_{u}^{c},\label{pro15a}\\
\mbox{s.t.}~
&\|\mathbf{w}_{a}\|^{2}\leq P_{max}^{DL},&\label{pro15b}\\
&\mathbf{t}_{a}\in\mathcal{C}_{t},\mathbf{r}_{a}\in\mathcal{C}_{r},&\label{pro15c}\\
&\|\mathbf{z}_{a}\|^{2}=1,\|\mathbf{u}_{a,u}\|^{2}=1,&\label{pro15d}\\
&\sum_{u=1}^{U}p_{u}\leq P_{max}^{UL}&\label{pro15e}\\
&|\mathbf{u}_{\mathcal{A}}(n,1)|=1,&\label{pro15f}\\
&\Delta f_{\min}\leq \Delta f_{a,a^{\prime}}\leq \Delta f_{\max},&\label{pro15g}
\end{align}\label{pro15}%
\end{subequations}
where the weight parameter $\beta$ in the objective function controls the trade-off between sensing and communication rates. 
This optimization is subject to several system constraints. Specifically, constraint~(\ref{pro15b}) limits the downlink transmit power, where $P_{\max}^{DL}$ denotes the maximum allowable downlink transmit power. Constraint~(\ref{pro15c}) restricts the feasible movement region of the MAs. Constraint~(\ref{pro15d}) enforces the normalization of the receive beamforming vectors. Constraint~(\ref{pro15e}) imposes the uplink transmit power constraint, where $P_{\max}^{UL}$ represents the maximum uplink transmit power. Constraint~(\ref{pro15f}) ensures the constant modulus property of the CFO, and constraint~(\ref{pro15g}) defines the allowable CFO range, with $\Delta f_{\min}$ and $\Delta f_{\max}$ denoting the minimum and maximum CFO values, respectively.


\section{Proposed Solution}
This section focuses on solving the joint uplink and downlink optimization problem 
(\ref{pro15}). To address this, the original problem is first decoupled into two subproblems: a worst-case CFO optimization problem and a resource allocation problem aimed at maximizing the {WCSR expression in (\ref{pro15})}. This two subproblems are formulated as follows:
\begin{subequations}
\begin{align}
\min_{\mathbf{u}_{\mathcal{A},\Delta f_{a,a^{\prime}}}}&~\beta\gamma^{r}+(1-\beta)\sum\nolimits_{u=1}^{U}\gamma_{u}^{c},\label{pro16a}\\
\mbox{s.t.}~
&(\ref{pro15f}),(\ref{pro15g}),&\label{pro16b}
\end{align}\label{pro16}
\end{subequations}
and
\begin{subequations}
\begin{align}
\max_{\mathbf{z},\{p_{u}\},\{\mathbf{u}_{u}\},\atop\{\mathbf{w}_{a}\},\{\mathbf{t}_{a}\},\{\mathbf{r}_{a}\}}&~\beta\gamma^{r}+(1-\beta)\sum\nolimits_{u=1}^{U}\gamma_{u}^{c},\label{pro17a}\\
\mbox{s.t.}~
&(\ref{pro15b})-(\ref{pro15g}),&\label{pro17b}
\end{align}\label{pro17}
\end{subequations}

\subsection{Proposed solution to subproblem (\ref{pro16})}
For the worst-case robust optimization in (\ref{pro16}), we find that $\gamma^{r}$ and $\gamma^{c}_{u}$ are all fractional form. Therefore, to deal with the fractional form in $\gamma^{r}$ and $\gamma^{c}_{u}$, we propose a fractional programming (FP) method. Specifically, introducing an auxiliary variable $\eta$ to achieve the quadratic transformation. $\gamma^{r}$ and $\gamma^{c}_{u}$ can be respectively rewritten as
\begin{align}
&\mathcal{F}_{r}(\mathbf{u}_{\mathcal{A}})=2\eta\sqrt{\mathbf{z}^{H}\bar{\mathbf{B}}(\mathbf{w}_{\mathcal{A}},\mathbf{t}_{\mathcal{A}},\mathbf{r}_{\mathcal{A}})\bar{\mathbf{B}}^{H}(\mathbf{w}_{\mathcal{A}},\mathbf{t}_{\mathcal{A}},\mathbf{r}_{\mathcal{A}})\mathbf{z}}\nonumber\\
&-\eta^{2}\left(\mathbf{u}_{\mathcal{A}}^{H}\tilde{\mathbf{D}}^{H}(\mathbf{r}_{\mathcal{A}},\mathbf{t}_{\mathcal{A}},\mathbf{w}_{\mathcal{A}})\mathbf{z}\mathbf{z}^{H}\tilde{\mathbf{D}}(\mathbf{r}_{\mathcal{A}},\mathbf{t}_{\mathcal{A}},\mathbf{w}_{\mathcal{A}})\mathbf{u}_{\mathcal{A}}\right.\nonumber\\
&\left.\mathbf{u}_{\mathcal{A}}^{H}\tilde{\mathbf{E}}^{H}(\mathbf{r}_{\mathcal{A}},\mathbf{t}_{\mathcal{A}},\mathbf{w}_{\mathcal{A}})\mathbf{z}\mathbf{z}^{H}\tilde{\mathbf{E}}(\mathbf{r}_{\mathcal{A}},\mathbf{t}_{\mathcal{A}},\mathbf{w}_{\mathcal{A}})\mathbf{u}_{\mathcal{A}}+\bar{c}_{1}\right)\nonumber\\
&=2\eta\sqrt{c_{1}}-\eta^{2}\left(\mathbf{u}_{\mathcal{A}}^{H}\tilde{\mathbf{C}}(\mathbf{r}_{\mathcal{A}},\mathbf{t}_{\mathcal{A}},\mathbf{w}_{\mathcal{A}})\mathbf{u}_{\mathcal{A}}+\bar{c}_{1}\right)\label{pro18}
\end{align}
and
\begin{align}
&\mathcal{F}_{c}(\mathbf{u}_{\mathcal{A}})=2\bar{\eta}\sqrt{p_{u}^{2}S^{2}\hat{\mathbf{u}}_{u}^{H}\mathbf{h}_{u}\mathbf{h}_{u}^{H}\hat{\mathbf{u}}_{u}}-\bar{\eta}^{2}\left(
\mathbf{u}_{\mathcal{A}}^{H}\bar{\mathbf{D}}^{H}(\mathbf{w}_{\mathcal{A}},\mathbf{t}_{\mathcal{A}},\mathbf{r}_{\mathcal{A}})\right.\nonumber\\
&\left.\hat{\mathbf{u}}_{u}\hat{\mathbf{u}}_{u}^{H}\bar{\mathbf{D}}^{H}(\mathbf{w}_{\mathcal{A}},\mathbf{t}_{\mathcal{A}},\mathbf{r}_{\mathcal{A}})\mathbf{u}_{\mathcal{A}}+\mathbf{u}_{\mathcal{A}}^{H}\tilde{\mathbf{E}}^{H}(\mathbf{w}_{\mathcal{A}},\mathbf{t}_{\mathcal{A}},\mathbf{r}_{\mathcal{A}})\hat{\mathbf{u}}_{u}\hat{\mathbf{u}}_{u}^{H}\right.\nonumber\\
&\left.\tilde{\mathbf{E}}^{H}(\mathbf{w}_{\mathcal{A}},\mathbf{t}_{\mathcal{A}},\mathbf{r}_{\mathcal{A}})\mathbf{u}_{\mathcal{A}}+c_{2,u}
\right)\nonumber\\
&=2\bar{\eta}\sqrt{c_{2,u}}-\bar{\eta}^{2}\left(\mathbf{u}_{\mathcal{A}}^{H}\tilde{\mathbf{C}}(\mathbf{r}_{\mathcal{A}},\mathbf{t}_{\mathcal{A}},\mathbf{w}_{\mathcal{A}})\mathbf{u}_{\mathcal{A}}+\bar{c}_{2,u}\right),\label{pro19}
\end{align}    
where 
\begin{align}
&c_{1}=\mathbf{z}^{H}\bar{\mathbf{B}}(\mathbf{w}_{\mathcal{A}},\mathbf{t}_{\mathcal{A}},\mathbf{r}_{\mathcal{A}})\bar{\mathbf{B}}^{H}(\mathbf{w}_{\mathcal{A}},\mathbf{t}_{\mathcal{A}},\mathbf{r}_{\mathcal{A}})\mathbf{z},\nonumber\\
&c_{2,u}=p_{u}^{2}S^{2}\hat{\mathbf{u}}_{u}^{H}\mathbf{h}_{u}\mathbf{h}_{u}^{H}\hat{\mathbf{u}}_{u},\nonumber\\
&\tilde{\mathbf{C}}=\tilde{\mathbf{D}}(\mathbf{r}_{\mathcal{A}},\mathbf{t}_{\mathcal{A}},\mathbf{w}_{\mathcal{A}})+\tilde{\mathbf{E}}(\mathbf{r}_{\mathcal{A}},\mathbf{t}_{\mathcal{A}},\mathbf{w}_{\mathcal{A}}),\nonumber\\
&\bar{\mathbf{C}}=\bar{\mathbf{D}}(\mathbf{r}_{\mathcal{A}},\mathbf{t}_{\mathcal{A}},\mathbf{w}_{\mathcal{A}})+\bar{\mathbf{E}}(\mathbf{r}_{\mathcal{A}},\mathbf{t}_{\mathcal{A}},\mathbf{w}_{\mathcal{A}}),\nonumber\\
&\bar{c}_{1}=|\mathbf{z}^{H}\bar{\mathbf{B}}(\mathbf{w}_{\mathcal{A}},\mathbf{t}_{\mathcal{A}},\mathbf{r}_{\mathcal{A}})|^{2}+|\mathbf{z}^{H}\bar{\mathbf{D}}(\mathbf{w}_{\mathcal{A}},\mathbf{t}_{\mathcal{A}},\mathbf{r}_{\mathcal{A}})|^{2}\nonumber\\
&+|\mathbf{z}\bar{\mathbf{\Xi}}(\mathbf{r}_{\mathcal{A}})|^{2}+\sigma^{2}\|\mathbf{z}\|^{2},\nonumber\\
&\bar{c}_{2,u}=|\hat{\mathbf{u}}_{u}^{H}\bar{\mathbf{B}}(\mathbf{w}_{\mathcal{A}},\mathbf{t}_{\mathcal{A}},\mathbf{r}_{\mathcal{A}})|^{2}+|\hat{\mathbf{u}}_{u}^{H}\bar{\mathbf{D}}(\mathbf{w}_{\mathcal{A}},\mathbf{t}_{\mathcal{A}},\mathbf{r}_{\mathcal{A}})|^{2}+\nonumber\\
&\sum\nolimits_{u^{\prime}\neq u}^{U}p_{u^{\prime}}^{2}S^{2}|\hat{\mathbf{u}}_{u}^{H}\mathbf{h}_{u^{\prime}}|^{2}+\sigma^{2}\|\hat{\mathbf{u}}_{u}\|^{2}.
\end{align} 
Since $c_{1}$ and $c_{2,u}$ are independent of the CFO vector $\mathbf{u}_{\mathcal{A}}$, they are treated as constants in the analysis.
The optimal values of $\eta$ and $\bar{\eta}$ can be updated by the following
\begin{align}
&\eta^{*}=\frac{\sqrt{c_{1}}}{\mathbf{u}_{\mathcal{A}}^{H}\tilde{\mathbf{C}}(\mathbf{r}_{\mathcal{A}},\mathbf{t}_{\mathcal{A}},\mathbf{w}_{\mathcal{A}})\mathbf{u}_{\mathcal{A}}+\bar{c}_{1}},\nonumber\\
&\bar{\eta}^{*}=\frac{\sqrt{c_{2,u}}}{\mathbf{u}_{\mathcal{A}}^{H}\bar{\mathbf{C}}(\mathbf{r}_{\mathcal{A}},\mathbf{t}_{\mathcal{A}},\mathbf{w}_{\mathcal{A}})\mathbf{u}_{\mathcal{A}}+\bar{c}_{2,u}}.\label{pro20}
\end{align}    
Substituting {(\ref{pro20})} into (\ref{pro18}) and (\ref{pro19}) such that (\ref{pro16a}) is rewritten as
\begin{subequations}
\begin{align}
\min_{\mathbf{u}_{\mathcal{A}},\Delta f_{a,a^{\prime}}}&~\beta \frac{c_{1}}{\mathbf{u}_{\mathcal{A}}^{H}\tilde{\mathbf{C}}(\mathbf{r}_{\mathcal{A}},\mathbf{t}_{\mathcal{A}},\mathbf{w}_{\mathcal{A}})\mathbf{u}_{\mathcal{A}}+\bar{c}_{1}}+\nonumber\\
&(1-\beta)\sum\nolimits_{u=1}^{U}\frac{c_{2,u}}{\mathbf{u}_{\mathcal{A}}^{H}\mathbf{C}(\mathbf{r}_{\mathcal{A}},\mathbf{t}_{\mathcal{A}},\mathbf{w}_{\mathcal{A}})\mathbf{u}_{\mathcal{A}}+\bar{c}_{2,u}},\label{pro21a}\\
\mbox{s.t.}~
&|\mathbf{u}_{\mathcal{A}}(n,1)|=1,\label{pro21b}&\\
&\Delta f_{\min}\leq \Delta f_{a,a^{\prime}}\leq \Delta f_{\max},\label{pro21c}&
\end{align}\label{pro21}%
\end{subequations}
Problem $(\ref{pro21})$ is still non-convex, to address this problem, problem $(\ref{pro21})$ is further divided into two subproblems, which are solved using an alternating optimization algorithm.
To simplify the subsequent analysis, the variables within the parentheses of 
$\tilde{\mathbf{C}}(\mathbf{r}_{\mathcal{A}},\mathbf{t}_{\mathcal{A}},\mathbf{w}_{\mathcal{A}})$ and $\mathbf{C}(\mathbf{r}_{\mathcal{A}},\mathbf{t}_{\mathcal{A}},\mathbf{w}_{\mathcal{A}})$ are omitted, and we let
$\mathbf{u}_{\mathcal{A}}=\boldsymbol{\phi}$. Therefore, the optimization problem is reformulated as
\begin{subequations}
\begin{align}
\min_{\mathbf{u}_{\mathcal{A}},\boldsymbol{\phi}}&~\beta \frac{c_{1}}{\boldsymbol{\phi}^{H}\tilde{\mathbf{C}}\boldsymbol{\phi}+\bar{c}_{1}}+(1-\beta)\sum\nolimits_{u=1}^{U}\frac{c_{2,u}}{\boldsymbol{\phi}^{H}\tilde{\mathbf{C}}\boldsymbol{\phi}+\bar{c}_{2,u}},\label{pro22a}\\
\mbox{s.t.}~
&(\ref{pro15f}),(\ref{pro15g}),\label{pro22b}&\\
&\mathbf{u}_{\mathcal{A}}=\boldsymbol{\phi},\label{pro22c}&
\end{align}\label{pro22}%
\end{subequations}  
This problem can be equivalently transformed into the following maximization problem, and (\ref{pro22}) is rewritten as
\begin{subequations}
\begin{align}
\max_{\mathbf{u}_{\mathcal{A}},\boldsymbol{\phi}}&~\beta \frac{\boldsymbol{\phi}^{H}\tilde{\mathbf{C}}\boldsymbol{\phi}+\bar{c}_{1}}{c_{1}}+(1-\beta)\sum\nolimits_{u=1}^{U}\frac{\boldsymbol{\phi}^{H}\tilde{\mathbf{C}}\boldsymbol{\phi}+\bar{c}_{2,u}}{c_{2,u}},\label{pro23a}\\
\mbox{s.t.}~
&(\ref{pro15f}),(\ref{pro15g}),(\ref{pro22c}),\label{pro23b}&
\end{align}
\end{subequations}  
According to the standard approach in constrained optimization, we introduce a penalty term to incorporate the constraint into the objective function. Based on this, we reformulate the original problem as
\begin{subequations}
\begin{align}
\max_{\mathbf{u}_{\mathcal{A}},\boldsymbol{\phi}}&~\beta \frac{\boldsymbol{\phi}^{H}\tilde{\mathbf{C}}\boldsymbol{\phi}+\bar{c}_{1}}{c_{1}}+(1-\beta)\sum\nolimits_{u=1}^{U}\frac{\boldsymbol{\phi}^{H}\tilde{\mathbf{C}}\boldsymbol{\phi}+\bar{c}_{2,u}}{c_{2,u}}\nonumber\\
&+\lambda\|\mathbf{u}_{\mathcal{A}}-\boldsymbol{\phi}\|^{2},\label{pro24a}\\
\mbox{s.t.}~
&(\ref{pro15f}),(\ref{pro15g}),\label{pro24b}&
\end{align}
\end{subequations}   
in which
$\lambda$ is a positive constant. Based on the alternating optimization algorithm proposed in \cite{b19}, the original problem is further decomposed into the following two subproblems.
\begin{subequations}
\begin{align}
\max_{\mathbf{u}_{\mathcal{A}},\boldsymbol{\phi}}&~\beta \frac{\boldsymbol{\phi}^{H}\tilde{\mathbf{C}}\boldsymbol{\phi}+\bar{c}_{1}}{c_{1}}+(1-\beta)\sum\nolimits_{u=1}^{U}\frac{\boldsymbol{\phi}^{H}\tilde{\mathbf{C}}\boldsymbol{\phi}+\bar{c}_{2,u}}{c_{2,u}}+\nonumber\\
&\lambda\|\mathbf{u}_{\mathcal{A}}-\boldsymbol{\phi}\|^{2},\label{pro25a}\\
\mbox{s.t.}~
&(\ref{pro15f}),\label{pro25b}&
\end{align}\label{pro25}
\end{subequations}  
and
\begin{subequations}
\begin{align}
\min_{\mathbf{u}_{\mathcal{A}}}&~\|\mathbf{u}_{\mathcal{A}}-\boldsymbol{\phi}\|^{2},\label{pro26a}\\
\mbox{s.t.}~
&(\ref{pro15g}),\label{pro26b}&
\end{align}\label{pro26}
\end{subequations}  

\textbf{Manifold optimisation (MO) for Problem (\ref{pro25}):} According to the definition of manifolds provided in \cite{b34,b35}, the TS error vector 
$\boldsymbol{\phi}=[\boldsymbol{\phi}(1),\ldots,\boldsymbol{\phi}(ABS)]^{T}$ is regarded as an embedded submanifold of the Euclidean space 
$\mathbb{C}^{ABS\times 1}$, where the TS error vector is represented by 
\begin{align}
\mathcal{O}=\{\boldsymbol{\phi}\in\mathbb{C}^{ABS\times 1}:|\boldsymbol{\phi}(i_{1})|=1,~i_{1}=1,\ldots,ABS\},\label{pro27}
\end{align}
in which the vector $\boldsymbol{\phi}(i_{1})$ is the 
$i_{1}$th element of $\boldsymbol{\phi}$. In general, {the} manifold 
$\mathcal{O}$ is not as naturally suited for optimization as Euclidean or vector spaces. To address this limitation, a Riemannian manifold is introduced, characterized by an inner product that varies smoothly across the tangent spaces at each point. Similar to how the derivative of a complex-valued function provides a local linear approximation, the tangent space 
$\mathcal{T}_{\boldsymbol{\phi}}\mathcal{O}$  at point 
$\boldsymbol{\phi}$ serves as a local vector space approximation of the manifold 
$\mathcal{O}$. Moreover, the presence of an inner product enables the definition of various geometric concepts on the manifold. By further treating 
$\mathbb{C}$ as a space 
$\mathbb{R}^{2}$ equipped with a canonical inner product, the Euclidean metric 
$\mathbb{C}$ on the complex plane can be accordingly defined as
\begin{align}
<\boldsymbol{\phi}(1),\boldsymbol{\phi}(2)>=\mathcal{R}[\boldsymbol{\phi}(1),\boldsymbol{\phi}^{*}(2)].\label{pro29}
\end{align}
Based on the definition of the inner product, the tangent vector of $\mathcal{O}$ can be defined accordingly. For a vector 
$\boldsymbol{\phi}$, if the inner product between each element in 
$\boldsymbol{\phi}$ and its corresponding element in 
$\boldsymbol{\phi}$ is zero, then 
$\boldsymbol{\phi}$ is said to be orthogonal to 
$\boldsymbol{\psi}$, i.e.,
\begin{align}
<\boldsymbol{\psi}(j),\boldsymbol{\phi}(j)>=0,\forall~j,\label{pro30}
\end{align}
in which the element 
$\boldsymbol{\psi}_{i}$ is the 
$s$th element of the set 
$\boldsymbol{\psi}$. By interpreting each complex-valued element of a vector as a vector in 
$\mathbb{R}^{2}$, 
$\boldsymbol{\psi}$ can be regarded as the tangent vector of 
$\mathcal{O}$ at point 
$\boldsymbol{\phi}$, provided that the following condition is satisfied
\begin{align}
\mathrm{Re}[\boldsymbol{\psi}\odot\boldsymbol{\phi}^{*}]=\mathbf{0}.\label{pro31}
\end{align}
The tangent space of the manifold 
$\mathcal{O}$ at point 
$\boldsymbol{\phi}$, denoted by 
$\mathcal{T}_{\boldsymbol{\phi}}\mathcal{O}$, is the set of all tangent vectors of 
$\mathcal{O}$ at point 
$\boldsymbol{\phi}$. Accordingly, the tangent space of 
$\boldsymbol{\phi}$ is expressed as
\begin{align}
\mathcal{T}_{\boldsymbol{\phi}}\mathcal{O}=\{\boldsymbol{\psi}\in\mathbb{C}^{S\times 1}:\mathcal{R}[\boldsymbol{\psi}\odot\boldsymbol{\phi}^{*}]=\mathbf{0}\}.\label{pro32}
\end{align}
To solve {(\ref{pro25})}, we employ the {Riemannian conjugate gradient (RCG)} method, an extension
of the traditional CG method to a Riemannian manifold. In
this approach, the update equations for the conjugate direction
$\mathbf{d}$
and the phase shift vector $\boldsymbol{\phi}$ at the $i$th iteration are
\begin{align}
\mathbf{d}_{i}=-\nabla_{\boldsymbol{\phi}}L(\boldsymbol{\phi}_{i})+\beta_{i}\mathcal{P}_{\mathcal{T}_{\boldsymbol{\phi}_{i}}\mathcal{O}}(\mathbf{d}_{i-1}), 
\end{align}
where $\beta_{i}=\frac{\|\nabla_{\boldsymbol{\phi}}L(\boldsymbol{\phi}_{i})\|^{2}}{\|\nabla_{\boldsymbol{\phi}}L(\boldsymbol{\phi}_{i-1})\|^{2}}$ is the Fletcher-Reeves CG parameter and $\alpha_{i}$
is the step size determined by the line search \cite{b36}. The update equations are repeated until $\boldsymbol{\phi}$ converges.
The projection of Euclidean gradient vector $\nabla_{\boldsymbol{\phi}} L(\boldsymbol{\phi})$ onto the tangent space $\mathcal{T}_{\boldsymbol{\phi}}$ {is a} unique vector $\mathcal{P}_{\mathcal{T}_{\boldsymbol{\phi}}\mathcal{O}}\left(\nabla_{\boldsymbol{\phi}} L(\boldsymbol{\phi})\right)$ that satisfies
\begin{align}
g(\nabla_{\boldsymbol{\phi}} L(\boldsymbol{\phi})-\mathcal{P}_{\mathcal{T}_{\boldsymbol{\phi}}\mathcal{O}}\left(\nabla_{\boldsymbol{\phi}} L(\boldsymbol{\phi})\right),\mathbf{q})=0,\forall~\mathbf{q}\in\mathcal{T}_{\boldsymbol{\phi}}\mathcal{O},
\end{align}
where $g:\mathcal{T}_{\boldsymbol{\phi}}\mathcal{O}\times\mathcal{T}_{\boldsymbol{\phi}}\mathcal{O}\rightarrow\mathbb{R}$ on $\mathcal{T}_{\boldsymbol{\phi}}\mathcal{O}$ at each point $\boldsymbol{\phi}$ is a smoothly varying inner product.
Owing to the local {homomorphism} between neighborhoods on a manifold and Euclidean space\cite{b36}, optimization algorithms originally developed for Euclidean spaces can be locally extended to Riemannian manifolds. By leveraging the vector space structure provided by the tangent space, conventional line search methods can be effectively applied within this framework. Based on this principle, this paper proposes a Riemannian MO-based gradient algorithm for TS error vector $\boldsymbol{\phi}$.
Based on (\ref{pro15}), we define the cost function as 
\begin{align}
&L(\boldsymbol{\phi})=\beta \frac{\boldsymbol{\phi}^{H}\tilde{\mathbf{C}}\boldsymbol{\phi}+\bar{c}_{1}}{c_{1}}+(1-\beta)\sum\nolimits_{u=1}^{U}\frac{\boldsymbol{\phi}^{H}\tilde{\mathbf{C}}\boldsymbol{\phi}+\bar{c}_{2,u}}{c_{2,u}}\nonumber\\
&+\lambda\|\mathbf{u}_{\mathcal{A}}-\boldsymbol{\phi}\|^{2}.\label{pro33}
\end{align}
To minimize the {Cramer-Rao lower bound} at each AP, it is necessary to determine the direction of steepest descent of the cost function within the current tangent space. Under the Riemannian optimization framework, the Euclidean gradient is first computed as 
\begin{align}
&\nabla_{\boldsymbol{\phi}} L(\boldsymbol{\phi}) =
- \beta \cdot \tilde{\mathbf{C}} \boldsymbol{\phi}/c_{1}
-(1-\beta) \sum_{u=1}^{U}  \tilde{\mathbf{C}}^{(k)} \boldsymbol{\phi}/c_{2,u}+\nonumber\\
&\lambda (\boldsymbol{\phi} - \mathbf{u}_{\mathcal{A}}),\label{pro34}
\end{align}
Then projected orthogonally onto the tangent space to obtain the Riemannian gradient, and the projection is given by
\begin{align}
\mathcal{P}_{\mathcal{T}_{\boldsymbol{\phi}}\mathcal{O}}\left(\nabla_{\boldsymbol{\phi}} L(\boldsymbol{\phi})\right)=\nabla_{\boldsymbol{\phi}} L(\boldsymbol{\phi})-\mathrm{R}\left\{\boldsymbol{\phi}^{*}\odot\nabla_{\boldsymbol{\phi}} L(\boldsymbol{\phi})\right\}\odot\boldsymbol{\phi}.\label{pro35}
\end{align}
If the concept of moving along a vector direction were as straightforward as in Euclidean space, point 
$\boldsymbol{\phi}$ would move within the tangent space 
$\mathcal{T}_{\boldsymbol{\phi}}\mathcal{O}$. On a manifold, this idea is generalized by the concept of a retraction, which enables movement along a tangent vector while ensuring that 
$\boldsymbol{\phi}$ remains on the manifold 
$\mathcal{O}$. Given a point 
$\boldsymbol{\phi}$ on 
$\mathcal{O}$, a step size 
$\alpha$, and a search direction 
$\mathbf{d}$, the new point is computed as follows
\begin{align}
\mathrm{Ret}_{\boldsymbol{\phi}}(\alpha\mathbf{d})=\left[\frac{\boldsymbol{\phi}(1)+\alpha d_{1}}{|\boldsymbol{\phi}(1)+\alpha d_{1}|},\ldots,\frac{\boldsymbol{\phi}(ABS)+\alpha d_{S}}{|\boldsymbol{\phi}(ABS)+\alpha d_{S}|}\right]^{T}.\label{pro_36}
\end{align}
By iteratively computing the gradient and updating the point on the manifold, we develop a conjugate gradient algorithm for TS error vector $\boldsymbol{\phi}$.

\textbf{CVX Optimization for Problem (\ref{pro26}):}After obtaining the TS error vector, the objective function in (\ref{pro26a}) is transformed into 
$\sum_{s=1}^{S}\sum_{a=1}^{A}\sum_{a^{\prime}=1}^{B}|\angle\boldsymbol{\phi}(aa^{\prime}s)+T_{sym}\Delta f_{a,a^{\prime}}|^{2}$, and the corresponding optimization problem for computing $\bar{\tau}_{a,b}$ is formulated as 
\begin{subequations}
\begin{align}
\min_{\Delta f_{a,a^{\prime}}}\sum_{s=1}^{S}\sum_{a=1}^{A}\sum_{a^{\prime}=1}^{B}&~|\angle\boldsymbol{\phi}(aa^{\prime}s)+T_{sym}\Delta f_{a,a^{\prime}}|^{2},\label{pro36a}\\
\mbox{s.t.}~
&\Delta f_{min}\leq\Delta f_{a,a^{\prime}}\leq\Delta f_{max},&\label{pro36b}
\end{align}\label{pro36}%
\end{subequations}
Since problem (\ref{pro36}) is convex, the optimization problem can be solved by using {CVX}\cite{b37}. The MO algorithm is summarized in \textbf{Algorithm~1}. The convergence analysis of MO algorithm is given in \textbf{Appendix~\ref{APP1}}.
\begin{algorithm}%
\caption{Proposed Algorithm for Problem (\ref{pro16})} \label{algo1}
\hspace*{0.02in}{\bf Initialize:}
$\Delta f_{a,a^{\prime}}^{(0)}$\\
\hspace*{0.02in}{\bf Repeat:}~$t=1,\ldots,T$.\\
\hspace*{0.02in}{Update $\eta^{*}$ and $\bar{\eta}^{*}$}\\
$\eta^{*}=\frac{\sqrt{c_{1}}}{\mathbf{u}_{\mathcal{A}}^{H}\tilde{\mathbf{C}}(\mathbf{r}_{\mathcal{A}},\mathbf{t}_{\mathcal{A}},\mathbf{w}_{\mathcal{A}})\mathbf{u}_{\mathcal{A}}+\bar{c}_{1}}$\\
$\bar{\eta}^{*}=\frac{\sqrt{c_{2,u}}}{\mathbf{u}_{\mathcal{A}}^{H}\bar{\mathbf{C}}(\mathbf{r}_{\mathcal{A}},\mathbf{t}_{\mathcal{A}},\mathbf{w}_{\mathcal{A}})\mathbf{u}_{\mathcal{A}}+\bar{c}_{2,u}}$\\
\hspace*{0.02in}{\bf while $\boldsymbol{\phi}$ does not converge do}\\
$\nabla_{\boldsymbol{\phi}} L(\boldsymbol{\phi}^{(t)})=\mathcal{P}_{\mathcal{T}_{\boldsymbol{\phi}^{(t)}}\mathcal{O}}(\nabla_{\boldsymbol{\phi}} L(\boldsymbol{\phi}^{(t)}))$\\
$\beta^{(t)}=\frac{\|\nabla_{\boldsymbol{\phi}}L(\boldsymbol{\phi}^{(t)})\|^{2}}{\|\nabla_{\boldsymbol{\phi}}L(\boldsymbol{\phi}^{(t-1)})\|^{2}}$\\
$\mathbf{d}^{(t)}=-\nabla_{\boldsymbol{\phi}}L(\boldsymbol{\phi}^{(t)})+\beta^{(t)}\mathcal{P}_{\mathcal{T}_{\boldsymbol{\phi}^{(t)}}\mathcal{O}}(\mathbf{d}^{(t-1)})$\\
Find a step size $\alpha^{(t)}$ based on Armijo Rule\cite{b36}\\
$\boldsymbol{\phi}^{(t+1)}=\mathrm{Ret}_{\boldsymbol{\phi}^{(t)}}(\alpha^{(t)}\mathbf{d}^{(t)})$\\
$t=t+1$\\
\hspace*{0.02in}{\bf end}\\
{Use} CVX to solve the problem in (\ref{pro36}).\\
\hspace*{0.02in}{\bf end}\\
\end{algorithm}

\subsection{Proposed solution to subproblem (\ref{pro17})}
In this section, we propose a MRL algorithm to optimize ${{\mathbf{w}_{a}},{\mathbf{t}_{a}},{\mathbf{r}_{a}}}$, aiming to maximize {the} WCSR. To introduce the proposed MRL algorithm, we begin by reviewing the fundamental concepts of DRL. Specifically, DRL algorithms {can be} trained in the CF-DFRC environment to select actions that maximize long-term rewards through trial-and-error interactions.
By collecting local observations from the environment, DRL helps reduce the computational overhead during training and accelerates convergence. Therefore, the Markov Decision Process (MDP) of DRL framework is defined as follows:
\begin{itemize}
    \item \textbf{State Space:} The current state $\mathbf{s}\in\mathcal{S}$ comprises the complete CSI at each AP, along with the MA positions [47], [48]. The state on the $t$th time is defined as 
    \begin{align}
    &\mathbf{s}_{t}=[\mathbf{h}_{u,a}^{(t-1)},\mathbf{g}_{u,a}^{(t-1)},\mathbf{H}_{a,a}^{(t-1)},\mathbf{G}_{a,b}^{(t-1)},\mathbf{H}_{T}^{(t-1)}],\label{pro37}
    \end{align}
where $\mathbf{h}_{u,a}^{(t-1)}$ and $\mathbf{g}_{u,a}^{(t-1)}$ represent the CSI of the $a$th AP to $u$th user at time $t-1$. $\mathbf{H}_{a,a}^{(t-1)}$,$\mathbf{G}_{a,b}^{(t-1)}$,$\mathbf{H}_{T}^{(t-1)}$ are the CSI of {the} SI channel, IAI channel and sensing channel.
\item \textbf{Action space:} The action space, denoted by $\mathcal{A}$, consists of a set of actions $a^{(t)}$, each of which triggers a transition from the current state $s^{(t)}$ to the next state $s^{(t+1)}$. Accordingly, each action is defined as  
\begin{align}
a^{(t)}=[\mathbf{w}_{a},\mathbf{t}_{a},\mathbf{r}_{a},\{p_{u}\}],\label{pro38}
\end{align} 
Within the action space, the MA position determines the feasible actions that the AP can take.  By incorporating the MA's location vector into both the action and state spaces, the reinforcement learning algorithm can effectively capture the relationship between AP control actions and the resulting changes in the propagation environment.
\item\textbf{Reward:} The reward is maximizing the WCSR which is expressed as
\begin{align}
r^{(t)}=\beta(\gamma^{r})^{(t)}+(1-\beta)\sum\nolimits_{k=1}^{K}(\gamma_{k}^{c})^{(t)},\label{pro39}
\end{align}
\end{itemize}
Based on the above definitions of DRL framework, we propose a MRL algorithm to solve Problem (\ref{pro17}) by jointly optimizing the transmit MA positions, receive MA positions, and transmit beamforming at the APs. The primary objective of the MRL algorithm is to learn an effective exploration policy that enhances the training efficiency of the actor network. The meta-policy is dynamically updated based on its success and failure rates. Within this framework, the MDP is modeled as a communication process between an exploration agent $(\pi_{e})$ and an exploitation agent $(\pi)$.  
$\pi$ is represented as 
\begin{align}
\pi=\delta(a^{*}(s^{t})-\mu(s^{t},\vartheta^{\pi_{e}}))  ,\label{pro40}  
\end{align}
where 
$a^{*}(s^{t})$ denotes the approximate optimal action for maximizing the Q-function, and 
$\mu(s^{t},\vartheta^{\pi_{e}})$ is a parameterized function. Since action 
$a^{*}(s^{t})$ includes both transmit beamforming and the MA positions, and is subject to power and position constraints, a projection operator is applied to $\mathbf{w}_{a}$, 
$\mathbf{t}_{a}$, and $\mathbf{r}_{a}$ after obtaining the optimal action 
$a^{*}(s^{t})$. This transformation ensures compliance with the maximum power and position constraints at the AP, as detailed below.
\begin{align}
&\prod(\mathbf{w}_{a})=\left\{
\begin{matrix}
\mathbf{w}_{a}&\mathrm{tr}(\mathbf{w}_{a}\mathbf{w}_{a}^{H})\leq P_{max}^{DL}\\
\frac{\sqrt{P_{max}^{DL}}}{\|\mathbf{w}_{a}\|_{F}}\mathbf{w}_{a}&\text{otherwise}
\end{matrix}\right.,\label{pro41}\\
&\prod(\mathbf{z}_{a})=\left\{
\begin{matrix}
\mathbf{z}_{a}&\mathrm{tr}(\mathbf{z}_{a}\mathbf{z}_{a}^{H})=1\\
\frac{\mathbf{z}_{a}}{\|\mathbf{z}_{a}\|_{F}}&\text{otherwise}
\end{matrix}\right.,\label{pro42}\\
&\prod(\mathbf{u}_{a,u})=\left\{
\begin{matrix}
\mathbf{u}_{a,u}&\mathrm{tr}(\mathbf{u}_{a,u}\mathbf{u}_{a,u}^{H})= 1\\
\frac{\mathbf{u}_{a,u}}{\|\mathbf{u}_{a,u}\|_{F}}&\text{otherwise}
\end{matrix}\right.,\\
&\prod(\mathbf{u}_{a,u})=\left\{
\begin{matrix}
p_{u}&\sum\nolimits_{u=1}^{U} p_{u}\leq P_{max}^{UL}\\
P_{max}^{UL}p_{u}/(\sum\nolimits_{u=1}^{U} p_{u})&\text{otherwise}
\end{matrix}\right.,\label{pro43}
\end{align}

The rollouts generated by executing action 
$\mathcal{B}_{0}$ are denoted as 
$\pi_{e}$, where $\mathcal{B}_{0}\in\mathcal{A}$. The meta-reward is then represented as 
\begin{align}
\mathcal{R}(\mathcal{B}_{0})=\mathcal{R}_{\pi^{\prime}}-\mathcal{R}_{\pi}  ,\label{pro44}  
\end{align}
The reward 
$\mathcal{R}_{\pi}$ is associated with the state transition 
$(\pi,\mathcal{B}_{0})\rightarrow\pi^{\prime}$. Within the MRL framework, this action can be interpreted as the generation of data $\mathcal{B}_{0}$
\begin{align}
\mathcal{L}(\pi_{e})=\mathbb{E}_{\mathcal{B}_{0}\sim\pi_{e}}[\mathcal{R}(\mathcal{B}_{0})]=\mathbb{E}_{\mathcal{B}_{0}\sim\pi_{e}}[\mathcal{R}_{D}(\pi,\mathcal{B}_{0})-\mathcal{R}_{\pi}],\label{pro45}
\end{align}
where the updated policy 
$\pi^{\prime}=P(\pi,\mathcal{B}_{0})$ is obtained through the DDPG. The true cumulative reward $r_{P(\pi,\mathcal{B}_{0})}$ is generated by policy $\pi^{\prime}$, while $r_{\pi}$ represents the true cumulative reward produced by policy $\pi$. Furthermore, the meta-reward is defined as the performance of policy $\mathcal{r}(\mathcal{B}_{0})$ evaluated under rollout $\mathcal{B}_{0}$. Accordingly, the exploration policy $\pi_{e}$ is parameterized by $\vartheta^{\pi_{e}}$, and the gradient of $\mathcal{L}(\pi_{e})$ with respect to $\vartheta^{\pi_{e}}$ is given by
\begin{align}
\nabla_{\vartheta^{\pi_{e}}}\mathcal{L}=\mathbb{E}_{\mathcal{B}\sim\pi_{e}}[\mathcal{r}(\mathcal{B}_{0})\nabla_{\vartheta^{\pi_{e}}}\log\mathcal{P}(\mathcal{B}_{0}|\pi_{e})],\label{pro46}
\end{align}
in which $\mathcal{P}(\mathcal{B}_{0}|\pi_{e})$ denotes the probability of transition tuples $\mathcal{B}_{0}=\{s^{t},a^{t},r^{t}\}_{t=1}^{T}$
given $\pi_{e}$ which is given by
\begin{align}
\mathcal{P}(\mathcal{B}_{0}|\pi_{e})=p(s^{0})\prod\limits_{t=0}^{n_{t}}\pi_{e}(a^{t}|s^{t})p(s^{t+1}|s^{t},a^{t}),\label{pro47}
\end{align}
where $p(s^{0})$ is the initial distribution, and $p(s^{t+1}|s^{t},a^{t})$ is the transition probability. Then, the gradient of $\mathcal{P}(\mathcal{B}_{0}|\pi_{e})$ with respect to $\vartheta^{\pi_{e}}$ is denoted as
\begin{align}
\nabla_{\vartheta^{\pi_{e}}}\log\mathcal{P}(\mathcal{F}_{0}|\pi_{e})=\sum\limits_{t=1}^{n_{t}}\nabla_{\vartheta^{\pi_{e}}}\log\pi_{e}(a^{t}|s^{t}).\label{pro48}
\end{align}
To estimate $r(\mathcal{B}_{0})$, the DDPG algorithm is executed using policy $\mathcal{F}_{0}$ to generate an updated policy for the actor network $\pi^{\prime}=D(\pi,\mathcal{F}_{0})$. Simulations are then initiated from the initial state $\pi^{\prime}$ to collect data, which includes information $\mathcal{F}_{1}$. This collected data is subsequently utilized to estimate the $\pi^{\prime}$-reward, leading to the computation of the meta-reward $\hat{r}_{\pi^{\prime}}$ as
\begin{align}
\hat{\mathcal{r}}(\mathcal{B}_{0})=\hat{r}_{\pi^{\prime}}-\hat{r}_{\pi}.\label{pro49}
\end{align}
Once $\mathcal{r}(\mathcal{B}_{0})$ has been estimated, we update the exploration policy $\pi_{e}$ by using the meta-policy gradient as defined in (14). The update rule is formulated as
\begin{align}
\vartheta^{\pi_{e}}=\vartheta^{\pi_{e}}+\eta \hat{\mathcal{R}}(\mathcal{F}_{0})\sum\limits_{t=1}^{T}\nabla_{\vartheta^{\pi_{e}}}\log\pi_{e}(a^{t}|s^{t}).\label{pro50}
\end{align}
Similarly, After updating the exploration policy, both $\mathcal{F}_{0}$ and $\mathcal{F}_{1}$ are incorporated into $r_{a}$ and subsequently exported to $\mathcal{r}_{a}=\mathcal{r}_{a}\cup\mathcal{F}_{0}\cup\mathcal{F}_{1}$. The participant policy $\pi$ is then updated, leading to $\pi=D(\pi,r_{a})$. The MRL algorithm is presented in \textbf{Algorithm~2}.
\begin{algorithm}%
\caption{Proposed Algorithm for Problem (\ref{pro17})} \label{algo2}
\hspace*{0.02in}{\bf Initialize:}
$\pi_{e}$, $\pi$, $\mathcal{B}_{1}$, $r_{a}$\\
\hspace*{0.02in}{\bf Repeat:}~$e=1,\ldots,E$.\\
Access $\mathbf{h}_{u,a}^{(t-1)}$,$\mathbf{g}_{u,a}^{(t-1)}$,$\mathbf{H}_{a,a}^{(t-1)}$,$\mathbf{G}_{a,b}^{(t-1)}$,$\mathbf{H}_{T}^{(t-1)}$\\
Randomly initial $\mathbf{h}_{u,a}^{(0)}$,$\mathbf{g}_{u,a}^{(0)}$,$\mathbf{H}_{a,a}^{(0)}$,$\mathbf{G}_{a,b}^{(0)}$,$\mathbf{H}_{T}^{(0)}$, and compute $s^{(0)}$\\
\hspace*{0.02in}{\bf Repeat:}~$t=1,\ldots,T$.\\
Generating $\mathcal{B}_{0}$ based on $\pi_{e}$, and using DDPG to update $\pi^{\prime}=d(\pi,\mathcal{B}_{0})$\\
Based on $\pi^{\prime}$, generate $\mathcal{B}_{1}$ and the reward of policy.\\
If the agent satisfies the 
all the constraints, the system
receives the reward according to (\ref{pro49}) and transits to the next state, otherwise, update the network parameters\\
Based on (\ref{pro50}), determine $\pi_{e}$ and $\mathcal{B}_{0}$ and $\mathcal{B}_{1}$ into $r_{a}$, i.e $r_{a}\longleftarrow r_{a}\cup\mathcal{B}_{0}\cup\mathcal{B}_{1}$\\
Based on DDPG, update $\pi\longleftarrow d(\pi,r_{a})$\\
Computing $\hat{\mathcal{B}}_{\pi}$\\
\hspace*{0.02in}{\bf end for}\\
\hspace*{0.02in}{\bf end for}\\
\end{algorithm}

\section{Numerical Results}\label{V}
\begin{figure}[htbp]
  \centering
  \includegraphics[width=0.3\textwidth, height=0.20\textwidth]{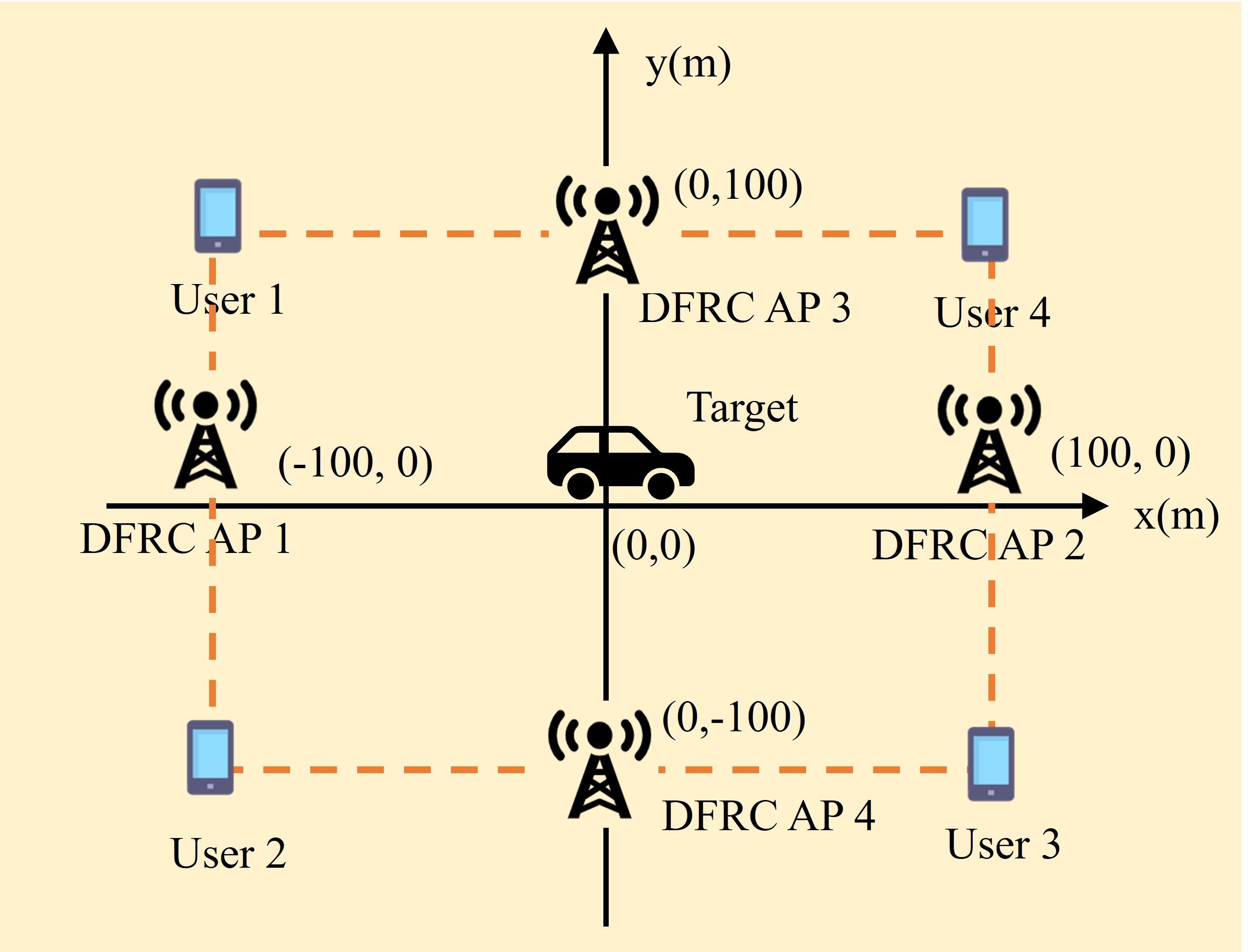}
  \captionsetup{justification=centering}
  \caption{Simulation setup of the MA-enabled C-DFRC system.}
\label{FIGURE3}
\end{figure}
In the simulation, the positions of the BSs, users, and targets are shown in \textbf{Fig}.\ref{FIGURE3}, where $A=4$, $U=4$, and {there is} one sensing target. Each BS is equipped with $N=8$ {transmit} MAs and $M = 4$ receive MAs. The channel path loss is modeled as $PL(d)=PL_{0}(d/d_{0})^{-\Omega}$, where $PL_{0}$ denotes the path loss factor, $d_{0}$ represents the reference distance, $d$ indicates the link distance, and $\Omega$ represents the path loss, with $PL_{0}=-30$~dB, $d_{0}=1$~m. The number of channel paths is $L_{u,a} =4$. The noise power level is $-120$~dBm. The minimum distance between MAs is set to $D= \lambda/2$. The range of the {transmit} {MAs} is $t_{\min}= -2\lambda$, $t_{\max}= 2\lambda$. The range for the receive {MAs} is $r_{min} = -2\lambda$, $r_{max} = 2\lambda$. The path loss exponent for the BS-user link is set to $2.8$, while the path loss exponent for the BS-target link is set to $2.2$. The target's RCS is given as $\alpha =0.5$. 
To demonstrate the performance of the proposed method, we compare its performance with three baselines. 
\begin{itemize}
    \item \textbf{Baseline 1}:The CFO is first estimated using the proposed robust MO {approach in algorithm 1}, followed by DRL-based optimization of the beamforming vector and MA position.
    \item \textbf{Baseline 2}:The method performs DRL-based joint optimization of all variables including CFO, beamforming, and MA location. 
    \item \textbf{Baseline 3}:The method utilizes the beamforming solution from FPAs as the initial point, and then applies the proposed robust algorithm together with SCA for further refinement.
\end{itemize}



\begin{figure}[htbp]
  \centering
  \includegraphics[scale=0.35]{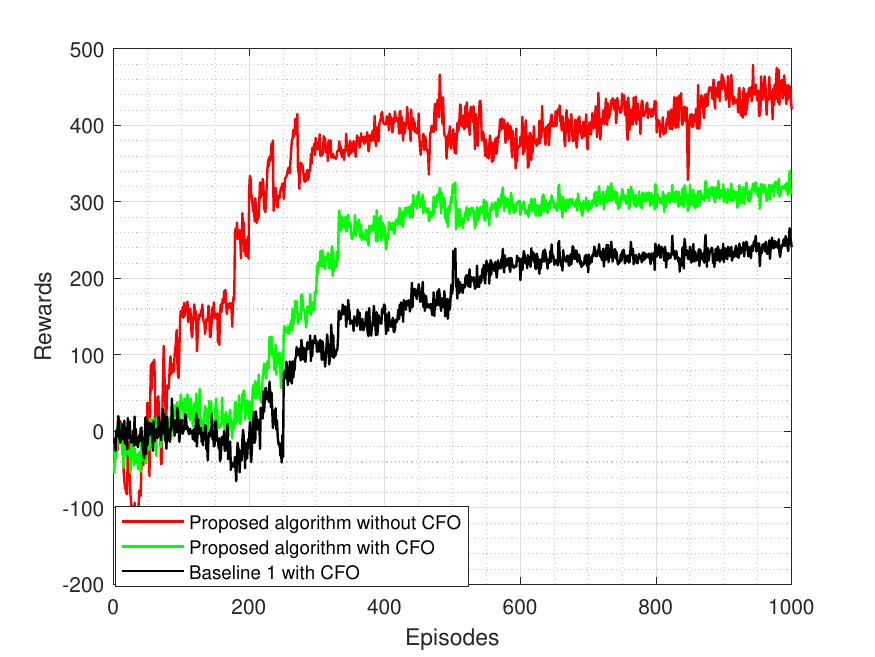}
  \captionsetup{justification=centering}
  \caption{Rewards versus episodes.}
\label{FIGURE31}
\end{figure}

Fig.\ref{FIGURE31} illustrates the convergence behavior of the proposed CFO-robust MRL algorithm in comparison with baseline 1 and 2. As the number of training episodes increases, all methods exhibit a general convergence trend, with the accumulated reward gradually stabilizing. Compared to the ideal scenario without CFO, our proposed method achieves performance that closely approximates the CFO-free case and demonstrates a faster convergence rate than baseline 1. This improvement is primarily attributed to the CFO-robust design incorporated into the MRL framework, which effectively integrates deterministic CFO features into the learning process. As a result, the proposed algorithm significantly mitigates the channel condition uncertainties caused by CFO, thereby improving overall sensing reliability.
Compared with conventional DRL {in baseline 1}, MRL accelerates convergence by learning a meta-policy that generalizes across multiple tasks. This enables rapid adaptation to new tasks with fewer iterations. Additionally, MRL reduces the trial-and-error overhead and benefits from more stable policy structures, thereby improving learning efficiency in dynamic environments.

\begin{figure}[htbp]
  \centering
  \includegraphics[scale=0.35]{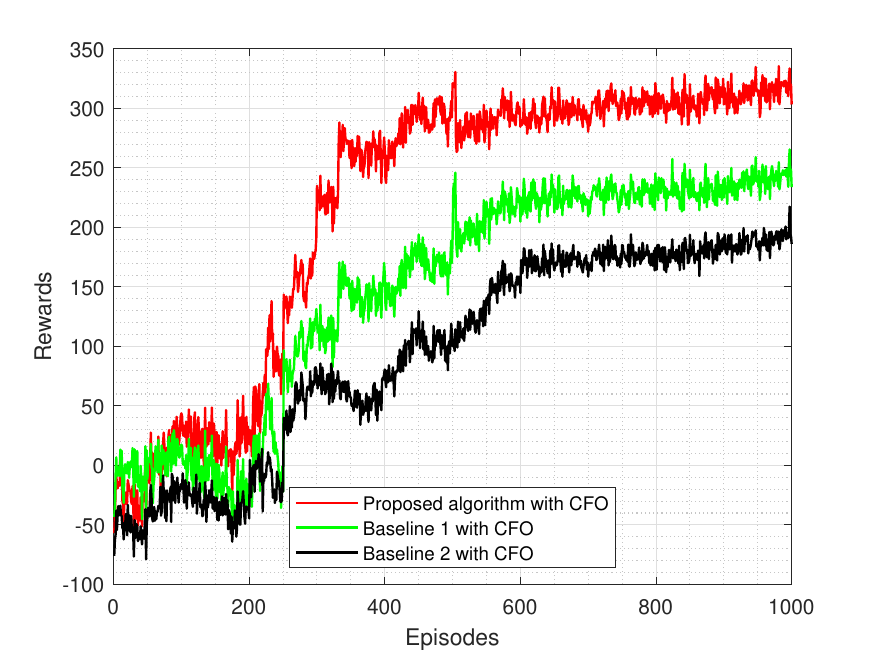}
  \captionsetup{justification=centering}
  \caption{Rewards versus episodes.}
\label{FIGURE4}
\end{figure}

Fig.\ref{FIGURE4} illustrates the convergence behavior of the proposed CFO-robust MRL algorithm in comparison with baseline 1 and 2. As the number of training episodes increases, all methods exhibit a general convergence trend, with rewards gradually stabilizing. However, the proposed approach consistently achieves faster convergence and attains a higher final reward. It is also evident that the {DRL} algorithm without CFO robustness performs the worst and converges the slowest. This is because baseline 2 {which does not approximate the CFO using algorithm 1}—{is} susceptible to noisy and distorted observations, which corrupt the reward signal and introduce significant variance during training. Moreover, completely neglecting the CFO effect expands the action space and makes its structure less coherent, as the agent must implicitly learn to compensate for CFO effects during training. This not only increases the burden of policy learning but also leads to instability and slower convergence due to the non-stationary nature of the environment.

\begin{figure}[htbp]
  \centering
  \includegraphics[scale=0.35]{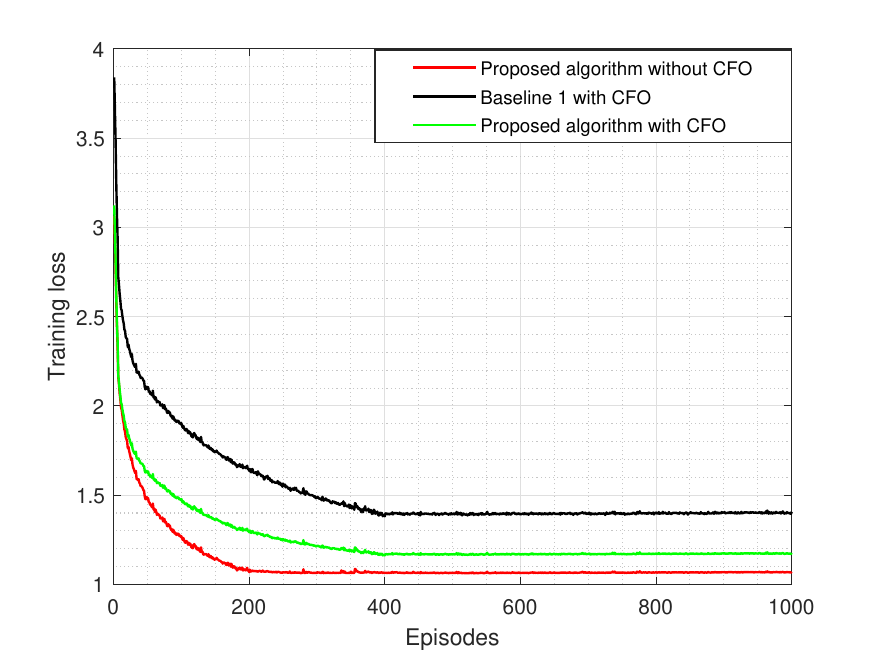}
  \captionsetup{justification=centering}
  \caption{Training loss versus episodes.}
\label{FIGURE5}
\end{figure}
As shown in Fig.~\ref{FIGURE5}, the training loss of all algorithms gradually decreases and eventually converges as the number of episodes increases. Compared with the baseline 1, the proposed algorithm achieves the lowest training loss and the fastest convergence rate. Specifically, in contrast to the MRL algorithm without CFO-robust design, our method effectively estimates the CFO, thereby mitigating the environment variations caused by CFO uncertainty and further reducing the action space. This contributes to a significantly faster convergence. Furthermore, compared with the conventional DRL algorithm, the MRL framework is capable of extracting shared policy priors across a distribution of related tasks, endowing the agent with better generalization capabilities. When encountering a new optimization task, the MRL-based agent can quickly adapt by leveraging the learned prior knowledge, requiring fewer interactions to attain near-optimal performance. In addition, the meta-learning mechanism facilitates more efficient exploration, which further enhances the overall learning efficiency and accelerates convergence.

\begin{figure}[htbp]
  \centering
  \includegraphics[scale=0.35]{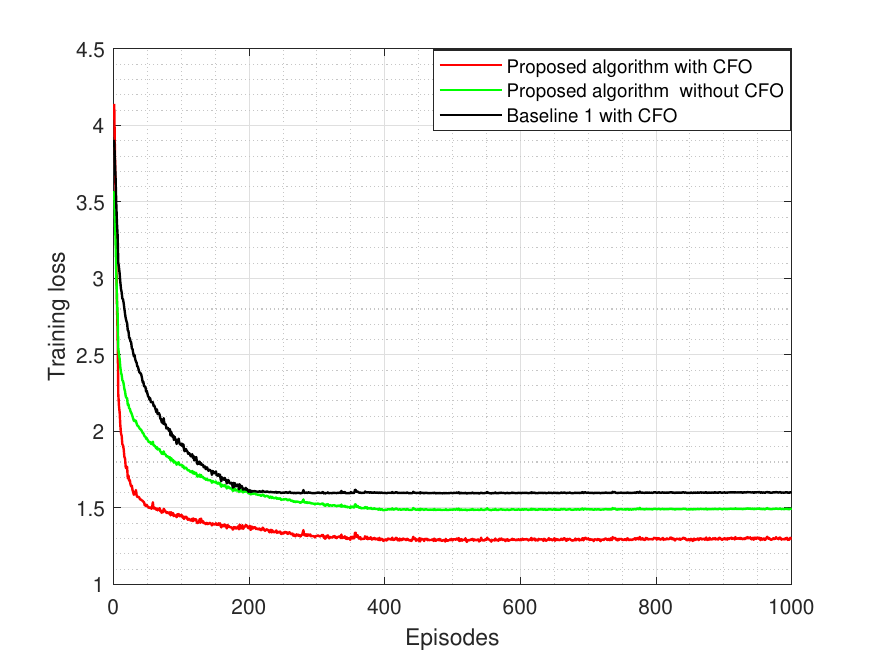}
  \captionsetup{justification=centering}
  \caption{Training loss versus episodes with $[-\lambda,\lambda]$.}
\label{FIGURE6}
\end{figure}
In Fig.~\ref{FIGURE6}, the movable region of the MA is reduced and set as region $[-\lambda,\lambda]$. As the number of training episodes increases, the training loss of all algorithms gradually decreases and converges. Compared with the other two baseline methods, the proposed algorithm achieves the lowest training loss and the fastest convergence. Compared with Fig.~\ref{FIGURE5}, all algorithms exhibit faster convergence in Fig.~\ref{FIGURE6}. This is attributed to the reduced movable region of the MA, which leads to a smaller action space. A smaller action space reduces the complexity of policy search and the output dimension of the policy network, thereby mitigating the uncertainty and exploration overhead during the learning process. 
However, compared with Fig.~\ref{FIGURE5}, the overall training loss in Fig.~\ref{FIGURE6} is higher for all algorithms. This is mainly due to the reduction in the MA’s movable range, which leads to a degradation in the DoFs of the system. The reduced DoF implies that the system’s capability to distinguish and serve spatial signal dimensions simultaneously is limited, thereby weakening its adaptability to complex channels in multi-user or multi-task scenarios. As a result, it becomes more difficult for the model to learn effective joint optimization strategies, leading to increased training loss and degraded performance.

\begin{figure}[htbp]
  \centering
  \includegraphics[scale=0.35]{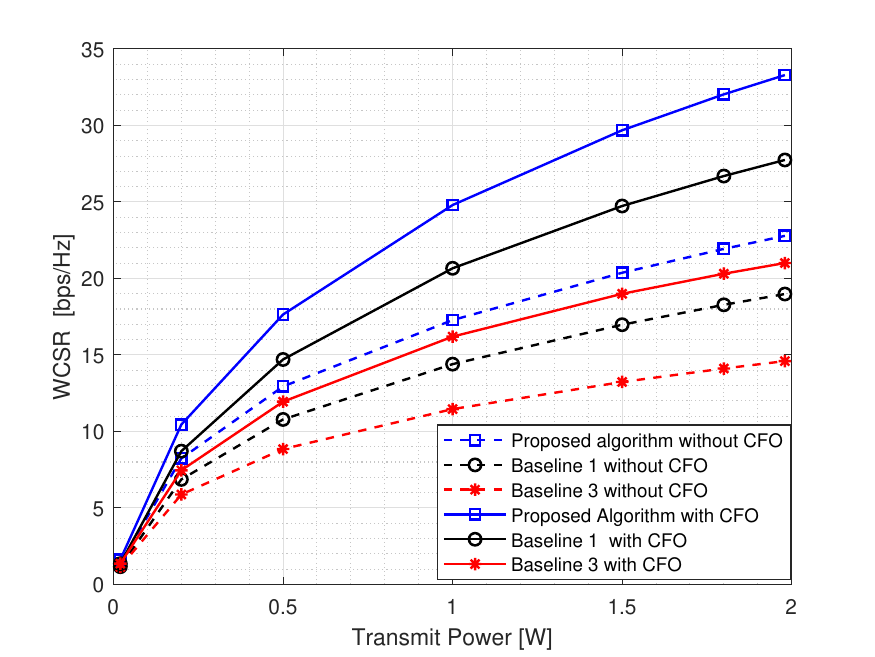}
  \captionsetup{justification=centering}
  \caption{WCSR versus episodes with Transmit power.}
\label{FIGURE7}
\end{figure}
In Fig.~\ref{FIGURE7}, as the transmit power increases, all three algorithms exhibit a consistent improvement in the WCSR, which aligns with theoretical expectations since higher power enhances both communication quality and sensing performance. In scenarios without CFO impairments, the proposed algorithm consistently outperforms the other two algorithms, achieving the highest WCSR. This superiority is attributed to its optimized policy design, which jointly addresses the coupling between communication and sensing objectives within a unified decision-making framework. In contrast to DRL-based methods that may lack effective policy training, the proposed method leverages meta-policy learning, enabling more efficient resource allocation and beamforming decisions. Furthermore, in scenarios with CFO impairments, the proposed algorithm still outperforms the other two, maintaining the highest WCSR. This robustness stems from the algorithm’s ability to mitigate the uncertainty introduced by CFO through its robust policy design.

\begin{figure}[htbp]
  \centering
  \includegraphics[scale=0.35]{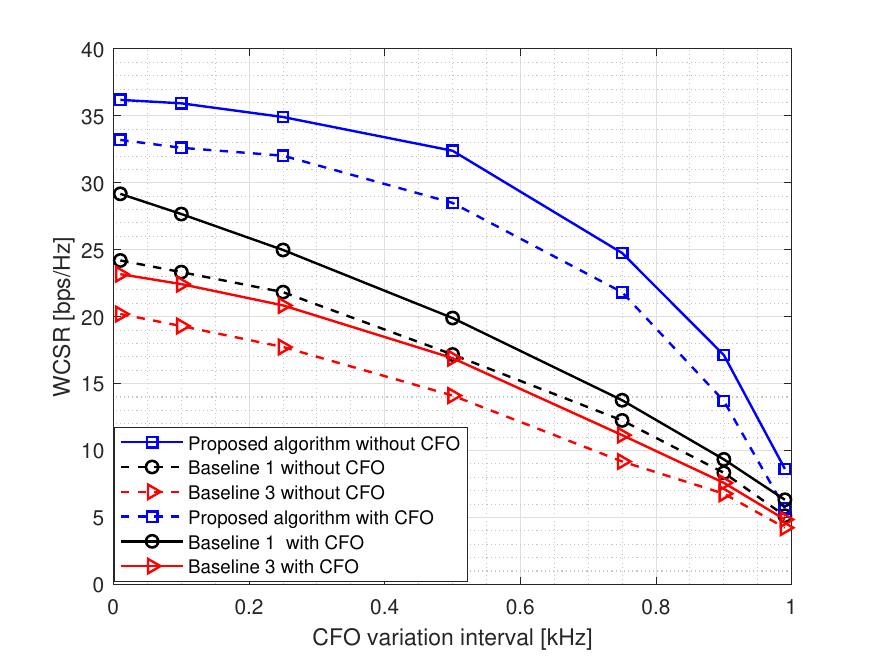}
  \captionsetup{justification=centering}
  \caption{WCSR versus CFO variation range.}
\label{FIGURE8}
\end{figure}
In Fig.~\ref{FIGURE8}, as the range of CFO increases, the WCSR of all algorithms gradually degrades. This degradation occurs because accurate beamforming and MA positions rely heavily on precise frequency synchronization to maintain coherent signal combining. A wider CFO range introduces more severe phase distortions, making it increasingly difficult for the system to align signals effectively—especially when the {users move} farther from the origin, where path loss and beam misalignment effects are exacerbated. Despite these challenges, the proposed algorithm exhibits superior robustness to {a wider range of} CFO impairments compared to the two baseline schemes.

\begin{figure}[htbp]
  \centering
  \includegraphics[scale=0.35]{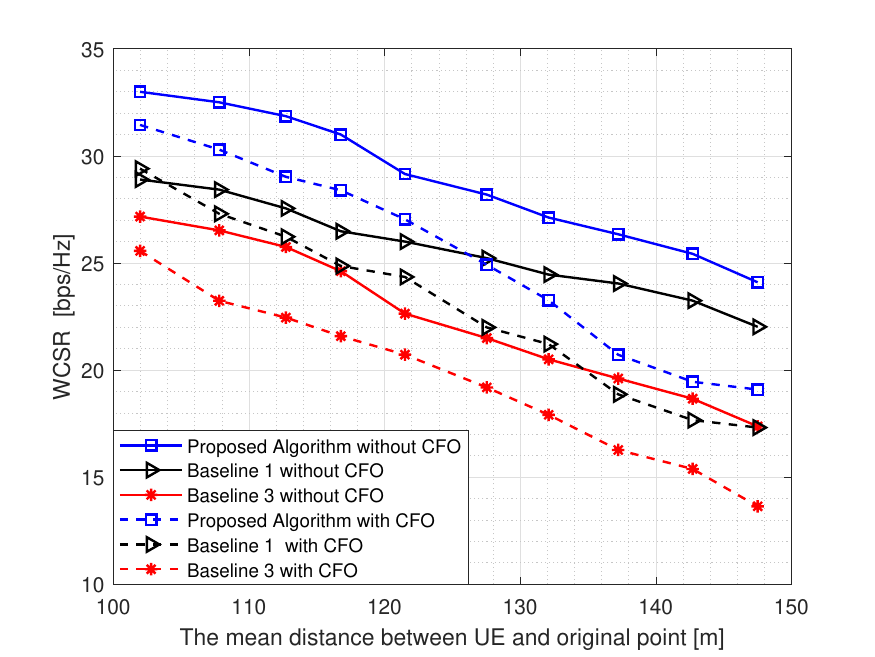}
  \captionsetup{justification=centering}
  \caption{WCSR versus the mean distance between {sensing target} and coordination original point.}
\label{FIGURE9}
\end{figure}
In Fig.~\ref{FIGURE9}, as the distance between the {sensing} target and the coordinate origin increases, the sensing rate of all three algorithms—including the proposed method and the two baseline schemes—gradually declines. This degradation is primarily attributed to increased path loss caused by longer propagation distances. Among them, the proposed algorithm experiences the slowest performance degradation, owing to its MRL framework, which enables rapid adaptation to varying target locations. Its meta-policy training architecture enhances the coordination between dynamic environmental changes and sensing strategies.
In contrast, the DRL-based scheme lacks such adaptability, leading to reduced beam control efficiency and degraded communication performance as the target moves farther from the origin. Furthermore, compared to the FPA-based scheme, all MA-enabled schemes outperform it, since the use of MAs provides higher DoF, allowing better alignment with LOS paths and more favorable channel conditions. This results in improved signal strength and enhanced communication reliability.

\section{Conclusion}\label{VI}
This paper proposed a robust optimization framework for wideband FD MA-aided CF-DFRC systems under CFO impairments. By jointly optimizing MA positions, beamforming vectors, and CFO parameters, the proposed approach effectively mitigates the adverse effects of CFO and enhances {the WCSR performance}. To tackle the resulting non-convex optimization problem, we developed a two-stage algorithm combining MO algorithm with PDD and MRL algorithm. The simulation results demonstrate that our proposed framework significantly outperforms conventional non-robust and DRL-based methods in terms of WCSR. Furthermore, the MA-aided system exhibits superior adaptability and robustness compared to FPA architectures, particularly in dynamic and CFO-impaired environments. These findings underscore the potential of MA-aided CF-DFRC systems as a promising solution for robust and efficient spectrum sharing in future 6G networks.

\begin{appendices}
\section{The proof of MO algorithm convergence}\label{APP1}
The boundedness of $L(\boldsymbol{\phi})$ can be inferred from its {form} (see (\ref{pro33})) and $\boldsymbol{\phi}$ satisfies  $\|\boldsymbol{\phi}\|_{2}=ABS$. Moreover,since $\nabla_{\boldsymbol{\phi}}L(\mathrm{Ret}_{\boldsymbol{\phi}}(\mathbf{d}))=\mathcal{P}_{\mathcal{T}_{\boldsymbol{\phi}}\mathcal{O}}(\nabla_{\boldsymbol{\phi}}L(\mathrm{Ret}_{\boldsymbol{\phi}}(\mathbf{d})))$ is a composite function of the projection operator $\mathcal{P}_{\mathcal{T}_{n}\mathcal{O}}(\cdot)$, the Euclidean gradient $\nabla_{\boldsymbol{\phi}}L(\cdot)$, and the retraction $\mathrm{retr}_{\boldsymbol{\phi}}(\cdot)$, we establish the Lipschitz continuity of $\nabla_{\boldsymbol{\phi}}L(\mathrm{Ret}_{\boldsymbol{\phi}}(\mathbf{d}))$ by demonstrating
the Lipschitz continuity of these three functions.

First, from the definition of the projection in (\ref{pro32}), we can verify that $\mathcal{P}_{\mathcal{T}_{\boldsymbol{\phi}}\mathcal{O}}$ is Lipschitz continuous as
\begin{align}
&\|\mathcal{P}_{\mathcal{T}_{\boldsymbol{\phi}}\mathcal{O}}(\mathbf{d}^{(t+1)})-\mathcal{P}_{\mathcal{T}_{\boldsymbol{\phi}}\mathcal{O}}(\mathbf{d}^{(t)})\|_{2}\leq \|\mathrm{Re}\{\boldsymbol{\phi}\odot(\mathbf{d}^{(t+1)}-\mathbf{d}^{(t)})\}\nonumber\\
&\odot\boldsymbol{\phi}\|_{2}+\|\mathbf{d}^{(t+1)}-\mathbf{d}^{(t)}\|_{2}\leq 2\|\mathbf{d}^{(t+1)}-\mathbf{d}^{(t)}\|_{2}.
\end{align}
Then, we can demonstrate the Lipschitz continuity of $L(\boldsymbol{\phi})$ as
\begin{align}
&|L(\boldsymbol{\phi}^{(t+1)})-L(\boldsymbol{\phi}^{(t)})|=\left|\beta\frac{(\boldsymbol{\phi}^{(t+1)})^{H}\tilde{\mathbf{C}}\boldsymbol{\phi}^{(t+1)}+\bar{c}_{1}}{c_{1}}+(1-\beta)\right.\nonumber\\
&\left.\sum_{u=1}^{U}\frac{(\boldsymbol{\phi}^{(t)})^{H}\tilde{\mathbf{C}}\boldsymbol{\phi}^{(t)}+\bar{c}_{1}}{c_{2,u}}+\lambda(\mathbf{u}_{\mathcal{A}}^{H}\mathbf{u}_{\mathcal{A}}+(\boldsymbol{\phi}^{(t)})^{H}\boldsymbol{\phi}^{(t)}\right.\nonumber\\
&\left.-2\mathrm{Re}(\mathbf{u}_{\mathcal{A}}^{H}\boldsymbol{\phi}^{(t)}))-\beta\frac{(\boldsymbol{\phi}^{(t+1)})^{H}\tilde{\mathbf{C}}\boldsymbol{\phi}^{(t+1)}+\bar{c}_{1}}{c_{1}}+(1-\beta)\sum_{u=1}^{U}\right.\nonumber\\
&\left.\frac{(\boldsymbol{\phi}^{(t+1)})^{H}\tilde{\mathbf{C}}\boldsymbol{\phi}^{(t+1)}+\bar{c}_{1}}{c_{2,u}}+\lambda(\mathbf{u}_{\mathcal{A}}^{H}\mathbf{u}_{\mathcal{A}}+(\boldsymbol{\phi}^{(t+1)})^{H}\boldsymbol{\phi}^{(t+1)}\right.\nonumber\\
&\left.-2\mathrm{Re}(\mathbf{u}_{\mathcal{A}}^{H}\boldsymbol{\phi}^{(t)}))\right|\leq\beta\left|\frac{(\boldsymbol{\phi}^{(t+1)})^{H}\tilde{\mathbf{C}}\boldsymbol{\phi}^{(t+1)}-(\boldsymbol{\phi}^{(t)})^{H}\tilde{\mathbf{C}}\boldsymbol{\phi}^{(t)}}{c_{1}}\right|\nonumber\\
&+(1-\beta)\sum\nolimits_{u=1}^{U}\left|\frac{(\boldsymbol{\phi}^{(t+1)})^{H}\tilde{\mathbf{C}}\boldsymbol{\phi}^{(t+1)}-(\boldsymbol{\phi}^{(t)})^{H}\tilde{\mathbf{C}}\boldsymbol{\phi}^{(t)}}{c_{2,u}}\right|\nonumber\\
&+2\lambda\left|\mathbf{u}_{\mathcal{A}}^{H}(\boldsymbol{\phi}^{(t+1)}-\boldsymbol{\phi}^{(t)})\right|\leq\beta\left|\lambda_{\max,\tilde{\mathbf{C}}}((\boldsymbol{\phi}^{(t+1)})^{H}\boldsymbol{\phi}^{(t+1)})\right.\nonumber\\
&\left.-\lambda_{\max,\tilde{\mathbf{C}}}((\boldsymbol{\phi}^{(t)})^{H}\boldsymbol{\phi}^{(t)})\right|+(1-\beta)\sum\nolimits_{u=1}^{U}\left|\lambda_{\max,\tilde{\mathbf{C}}}((\boldsymbol{\phi}^{(t+1)})^{H}\right.\nonumber\\
&\left.\boldsymbol{\phi}^{(t+1)})-\lambda_{\max,\tilde{\mathbf{C}}}((\boldsymbol{\phi}^{(t)})^{H}\boldsymbol{\phi}^{(t)})\right|+2\lambda|\mathrm{Re}(\mathbf{u}_{\mathcal{A}}^{H}(\boldsymbol{\phi}^{(t)}-\boldsymbol{\phi_{2}}))|\leq\nonumber\\
&\beta ABS|\lambda_{\max,\tilde{\mathbf{C}}}-\lambda_{\min,\tilde{\mathbf{C}}}|\nonumber\\
&+(1-\beta)ABS\sum\nolimits_{u=1}^{U}|\lambda_{\max,\tilde{\mathbf{C}}}-\lambda_{\min,\tilde{\mathbf{C}}}|2ABS\lambda\|\mathbf{u}_{\mathcal{A}}^{H}\|.
\end{align}    
The upper bound is a constant, thus, $L(\boldsymbol{\phi})$ is Lipschitz continuous.
Then, we continue to demonstrate the Lipschitz continuity of
$\mathrm{Ret}_{\boldsymbol{\phi}}(\mathbf{d})$. Let $\phi_{j}$ and $d_{j}$ be the jth elements of $\mathbf{\eta}$ and $\mathbf{d}$,
respectively, then we obtain
\begin{align}
&|\mathrm{Ret}_{\phi_{j}}(d_{j})-\mathrm{Ret}_{\phi_{j}}(0)|=\left|\frac{\phi_{j}+d_{j}}{|\phi_{j}+d_{j}|}-\phi_{j}\right|\nonumber\\
&=|\phi_{j}|\left|\frac{(1+d_{j}\phi_{j}^{-1})}{|\phi_{j}|(1+d_{j}\phi_{j}^{-1})}-1\right|=\left|\frac{1+d_{j}^{\prime}}{|1+d_{j}^{\prime}|}-1\right|
\end{align}    
where $d_{j}^{\prime}=d_{j}\phi_{j}^{-1}$. If $|d_{j}^{\prime}|=|d_{j}|\geq\frac{1}{2}$, then
\begin{align}
\left|\frac{1+d_{j}^{\prime}}{|1+d_{j}^{\prime}|}-1\right|\leq \left|\frac{1+d_{j}^{\prime}}{|1+d_{j}^{\prime}|}\right|+1=2\leq 4|d_{j}|
\end{align}    
When $|d_{j}^{\prime}|=|d_{j}|<\frac{1}{2}$, we have
\begin{align}
&\left|\frac{1+d_{j}^{\prime}}{|1+d_{j}^{\prime}|}-1\right|^{2}=\frac{(|1+d_{j}^{\prime}|-\mathrm{Re}\{1+d_{j}^{\prime}\})}{|1+d_{j}^{\prime}|}=\nonumber\\
&\frac{2\mathrm{Im}\{1+d_{j}^{\prime}\}}{|1+d_{j}^{\prime}|(|1+d_{j}^{\prime}|+\mathrm{Re}\{1+d_{j}^{\prime}\})}\leq\frac{2|d_{j}^{\prime}|^{2}}{|1+d_{j}^{\prime}|}\leq 8|d_{j}^{\prime}|
\end{align}
which shows the Lipschitz continuity of $\mathrm{Ret}_{\boldsymbol{\phi}}(\mathbf{d})$.
By combining the Lipschitz continuities of $\mathcal{P}_{\mathcal{T}_{\boldsymbol{\phi}}\mathcal{O}}$, $L(\boldsymbol{\phi})$ and $\mathrm{Ret}_{\boldsymbol{\phi}}$, we obtain the desired result. {Next}, we {analyze} the bound of $\nabla_{\boldsymbol{\phi}}L(\boldsymbol{\phi})$. According to\cite{b36}, since we use the Armijo rule to obtain {the} step size, $\alpha^{(t)}$ satisfies the strong Wolfe conditions, therefore, we have
\begin{align}
L(\mathrm{Ret}_{\boldsymbol{\phi}^{(t)}}(\alpha^{(t)}\mathbf{d}^{(t)}))\leq L(\boldsymbol{\phi}^{(t)})+\rho_{1}\alpha^{(t)}g(\nabla_{\boldsymbol{\phi}}L(\boldsymbol{\phi}^{(t)}),\mathbf{d}^{(t)}),\nonumber\\
|g(\nabla_{\boldsymbol{\phi}}L(\mathrm{Ret}{(\alpha^{(t)}\mathbf{d}^{(t)}})),\mathbf{d}^{(t)})|\leq -\rho_{2}g(\nabla_{\boldsymbol{\phi}}L(\boldsymbol{\phi}^{(t)}),\mathbf{d}^{(t)}),\label{app64}
\end{align}
where $0\leq\rho_{1}\leq\rho_{2}\leq 1$.
Finally, {we} {can rewrite} the second Wolfe condition (\ref{app64}) as
\begin{align}
g(\nabla_{\boldsymbol{\phi}}L(\mathrm{Ret}_{\boldsymbol{\phi}^{(t)}}(\alpha^{(t)}\mathbf{d}^{(t)})),\mathbf{d}^{(t)})\geq \rho_{2}g(\nabla_{\boldsymbol{\phi}}L(\boldsymbol{\phi}^{(t)}),\mathbf{d}^{(t)})
\end{align}
From this, we derive
\begin{align}
&(\rho_{2}-1)g(\nabla_{\boldsymbol{\phi}}L(\boldsymbol{\phi}^{(t)}),\mathbf{d}^{(t)})\leq \|g(\nabla_{\boldsymbol{\phi}}L(\mathrm{Ret}(\alpha^{(t)}\mathbf{d}^{(t)}))-\nonumber\\
&g(\nabla_{\boldsymbol{\phi}}L(\mathrm{Ret}(\mathbf{0}))\|_{2}\|\mathbf{d}^{(t)}\|_{2}
\end{align}
Given the Lipschitz continuity of the Riemannian gradient (i.e., $\|\nabla_{\boldsymbol{\phi}}L(\mathrm{Ret}(\alpha^{(t)}\mathbf{d}^{(t)}))-\mathrm{Ret}(\mathbf{0})\|_{2}\leq\alpha^{(t)}K\|\mathbf{d}^{(t)}\|_{2}$), we find
\begin{align}
(\rho_{2}-1)g(\nabla_{\boldsymbol{\phi}}L(\boldsymbol{\phi}^{(t)}),\mathbf{d}^{(t)})\leq \alpha^{(t)}K\|\mathbf{d}^{(t)}\|_{2}^{2}
\end{align}
Since $g(\nabla_{\boldsymbol{\phi}}L(\boldsymbol{\phi}^{(t)}),\mathbf{d}^{(t)})<0$, we have
\begin{align}
\rho_{1}\alpha^{(t)}g(\nabla_{\boldsymbol{\phi}}L(\boldsymbol{\phi}^{(t)}),\mathbf{d}^{(t)})\leq \rho_{1}\frac{\rho_{2}-1}{K}\frac{g(\nabla_{\boldsymbol{\phi}}L(\boldsymbol{\phi}^{(t)}),\mathbf{d}^{(t)})^{2}}{\|\mathbf{d}\|_{2}^{2}},\label{app68}
\end{align}
By combining (\ref{app68}) with the first Wolfe condition (\ref{app64}) where
$L(\boldsymbol{\phi}^{(t+1)})=L(\mathrm{Ret}_{\boldsymbol{\phi}^{(t)}}(\alpha^{(t)}\mathbf{d}^{(t)}))$, we obtain
\begin{align}
L(\boldsymbol{\phi}^{(t+1)})\leq L_(\boldsymbol{\phi}^{(t)})-c\cos^{2}\theta^{(t)}\|\nabla_{\boldsymbol{\phi}}L(\boldsymbol{\phi}^{(t)})\|_{2}^{2}
\end{align}
where $c=\rho_{1}(1-\rho_{2})/K>0$. Summing these expressions
over $j=1,2,\ldots,i$ yields:
\begin{align}
L(\boldsymbol{\phi}^{(t+1)})\leq L_(\boldsymbol{\phi}^{(t)})-c\sum\nolimits_{j=1}^{i}\cos^{2}\theta^{(t)}\|\nabla_{\boldsymbol{\phi}}L(\boldsymbol{\phi}^{(t)})\|_{2}^{2}
\end{align}
Since $L(\boldsymbol{\phi})$ is bounded below, there exists $\beta>0$ such that $L(\boldsymbol{\phi}^{(t)})-L(\boldsymbol{\phi}^{(t+1)})<\beta$ for all i. Consequently, we have
\begin{align}
c\sum\nolimits_{j=1}^{i}\cos^{2}\theta^{(t)}\|\nabla_{\boldsymbol{\phi}}L(\boldsymbol{\phi}^{(t)})\|_{2}^{2}\leq \beta
\end{align}
There, the gradient of $L(\boldsymbol{\phi})$ is bounded. In summary, since $\mathrm{Ret}_{\boldsymbol{\phi}}(\mathbf{d})$, 
$L(\boldsymbol{\phi})$, 
$\mathcal{P}_{\mathcal{T}_{\boldsymbol{\phi}}\mathcal{O}}$, and 
$\nabla_{\boldsymbol{\phi}}L(\boldsymbol{\phi})$ are all bounded, the proposed MO algorithm is guaranteed to converge. This completes the proof.
\end{appendices}

\end{document}